\documentclass[aip,reprint,pof]{revtex4-1}

\usepackage[T1]{fontenc}
\usepackage[utf8]{inputenc}
\usepackage[english]{babel}
\usepackage{graphics}
\usepackage{graphicx}
\usepackage{amsmath,amssymb,esint}
\usepackage{multirow}
\usepackage{subfig}
\usepackage{natbib}
\usepackage[usenames,x11names]{xcolor}
\usepackage[colorlinks,linkcolor=blue,citecolor=blue]{hyperref}
\usepackage{upgreek}
\usepackage{algorithm}
\usepackage{algpseudocode}

\usepackage{tikz}
\usetikzlibrary{shapes.geometric, arrows, intersections, through}
\usetikzlibrary{arrows.meta}
\usetikzlibrary{calc}
\tikzset{math3d/.style={z={(-0.65cm,-0.30cm)},y={(0cm,1cm)},x={(0.9cm,-0.15cm)}}}
\usetikzlibrary{shapes.geometric, arrows, intersections, through}
\usetikzlibrary{decorations.text, decorations.shapes, backgrounds}
\usetikzlibrary{arrows.meta}
\usetikzlibrary{colorbrewer}
\usetikzlibrary{patterns}

\renewcommand{\selectlanguage}[1]
{}

\begin{document}

\title{The Kirkwood-Bethe hypothesis for bubble dynamics, cavitation and underwater explosions}

\author{\vspace{0.7em}Fabian Denner}
\email[]{fabian.denner@polymtl.ca}
\affiliation{\mbox{Department of Mechanical Engineering, Polytechnique Montr\'eal, Montr\'eal, H3T 1J4, Québec,~Canada}}


\begin{abstract}
Pressure-driven bubble dynamics is a major topic of current research in fluid dynamics, driven by innovative medical therapies, sonochemistry, material treatments, and geophysical exploration. First proposed in 1942, the Kirkwood-Bethe hypothesis provides a simple means to close the equations that govern pressure-driven bubble dynamics as well as the resulting flow field and acoustic emissions in spherical symmetry. The models derived from the Kirkwood-Bethe hypothesis can be solved using standard numerical integration methods at a fraction of the computational cost required for fully resolved simulations. Here, the theoretical foundation of the Kirkwood-Bethe hypothesis and contemporary models derived from it are gathered and reviewed, as well as generalized to account for spherically symmetric, cylindrically symmetric, and planar one-dimensional domains. In addition, the underpinning assumptions are clarified and new results that scrutinize the predictive capabilities of the Kirkwood-Bethe hypothesis with respect to the complex acoustic impedance experienced by curved acoustic waves and the formation of shock waves are presented. Although the Kirkwood-Bethe hypothesis is built upon simplifying assumptions and lacks some basic acoustic properties, models derived from it are able to provide accurate predictions under the specific conditions associated with pressure-driven bubble dynamics, cavitation and underwater explosions. \vspace{0.3em}

\noindent 
\textsl{\small Copyright (2024) Author(s). This article is distributed under a Creative Commons Attribution (CC BY) License.}
\end{abstract}


\maketitle 

\vspace{-1em}
\section{Introduction}

Pressure-driven bubble dynamics are encountered and play an important role in countless engineering applications and natural phenomena \citep{Plesset1977,Brennen1995,Lauterborn2010}. The formation and collapse of vapor bubbles (\textit{cf.}~hydrodynamic cavitation) are well known to cause material erosion of propellers and hydro turbines \citep{Blake1987,Reuter2022,Dular2023}, and the related acoustic emissions are actively studied and monitored \textit{in situ} \citep{Johansen2017,Gaisser2023,Haskell2023}. However, the past decades have seen a paradigm shift in how pressure-driven bubble dynamics and, especially, cavitation are studied, away from treating cavitation solely as a hazard, towards leveraging its mechanical and thermal effects. This has spawned innovative new applications, such as contrast-enhanced medical imaging \citep{Wan2015}, intensifying chemical reactions \citep{Meroni2022}, breaking down kidney and urinary stones \citep{Maeda2018a, Sapozhnikov2021}, treating tumors \citep{Yeats2023}, using cavitation as a sonic scalpel for precision surgery \citep{Vogel1986,Lee2017b}, targeted drug delivery by ultrasound-driven micro- and nanobubbles \citep{Shakya2024}, 3D printing with sound \citep{Habibi2022}, actuating smart materials \citep{Athanassiadis2022}, and enhancing bioadhesion \citep{Ma2022}. The energy focused by a strong bubble collapse can be utilized to pierce biological cells \citep{Ohl2006, Helfield2016}, to destruct bacteria for wastewater treatment \citep{Sarc2018}, to clean surfaces and membranes \citep{Chahine2016,Reuter2017a}, to {synthesize} nanoparticles and nanostructured materials \citep{Barcikowski2019}, and to manipulate hardened materials, such as metals \citep{Soyama2022} and glass \citep{Gutierrez-Hernandez2023}. Ferns use cavitation-driven catapults to eject their spores \citep{Llorens2016} and pistol shrimps use the shock waves generated by collapsing bubbles to stun their prey \citep{Koukouvinis2017}. The pressure and shock waves of underwater explosions are studied for defense purposes as a major component of marine warfare \citep{Cole1948,Wu2020,Yu2023} and the pressure transients produced by the generation and collapse of underwater bubbles are used for seismic surveys in marine and geophysical exploration \citep{deGraaf2014,MacGillivray2019,Athanassiadis2019}. The advent of modern monitoring and control techniques for bubble dynamics, many of which now leverage machine learning algorithms, largely rely on acoustic signals to identify and classify bubble activity \citep{OReilly2012,Novell2020,Doan2022,Rom2023,Gaisser2023,Klapcsik2023,Mallik2024}. This, in turn, requires high-fidelity models of the complex bubble dynamics and nonlinear acoustics to classify the related acoustic signatures and to train machine learning algorithms.

The \textit{Kirkwood-Bethe hypothesis} \citep{Kirkwood1942} is an important tool for fundamental and theoretical research on pressure-driven bubble dynamics, as well as the related fluid dynamics and acoustics. It stipulates that in spherical symmetry the quantity 
\begin{equation}
    g = r \left(h-h_\infty + \frac{u^2}{2} \right),  \label{eq:kb}
\end{equation}
emitted at the gas-liquid interface of a gas cavity or bubble, is constant along outgoing characteristics and propagates  with speed $c+u$, where $r$ is the radial coordinate, $c$ is the liquid speed of sound, $u$ is the flow velocity, and $h-h_\infty$ is the difference in specific enthalpy between radial location $r$ and the ambient state denoted with subscript $\infty$. 
The power of the Kirkwood-Bethe hypothesis lies in the fact that it enables, in conjunction with a suitable equation of state, to close the set of equations governing the pressure-driven behavior of gas bubbles and the resulting flow of the surrounding liquid, such that mathematical models for the velocity and pressure generated by a cavitation bubble or an underwater explosion can be derived. 
Computer programs developed in the 1960s using models for pressure-driven bubble dynamics and acoustics based on the Kirkwood-Bethe hypothesis were also among the first open-source software tools in fluid dynamics \citep{Ivany1965,Lilliston1966}.
Despite the widespread availability of simulation tools that can predict two- and three-dimensional pressure-driven bubble dynamics in complex scenarios with increasing accuracy and robustness \citep{Denner2018b,Fuster2018,Schmidmayer2020a,Bryngelson2021,Saade2023}, the Kirkwood-Bethe hypothesis and the models derived from it, henceforth collectively referred to as \textit{KBH models}, still play an important role for the prediction of pressure-driven bubble phenomena. Predicting the dynamic behavior of collapsing bubbles and the pressure transients they produce remains challenging and KBH models have emerged as robust benchmarks for testing and validating these numerical methods \citep{Koch2016,Denner2020a,Gonzalez-Avila2021,Saade2023,Wang2024}. In addition, acoustic emissions can be modulated by different mechanisms \citep{Crighton1992,Christov2017,Denner2024} and, consequently, interpreting acoustic signals at a distance remains a complex task. To this end, the Kirkwood-Bethe hypothesis has seen a renewed interest over the past years as the basis for gaining an improved understanding of the interaction between acoustic emissions and fluid dynamics \citep{Lai2022,Liang2022,Denner2023,Wen2023a,Denner2024}, primarily in relation to pressure-driven bubble dynamics.

Although the Kirkwood-Bethe hypothesis and models are well established and widely used today, much of their theoretical foundation is scattered over many institutional reports \citep{Gilmore1952,Flynn1957,Hickling1963,Lilliston1966,Beyer1974,Flynn1978}, declassified military reports \citep{Rice1943,Rice1944,Whitham1953,Snay1966,Naugolnykh1971}, textbooks \citep{Cole1948, Cooper1958, Rozenberg1971, Lauterborn1980}, PhD theses \citep{Schneider1949,Ivany1965}, and papers \citep{Hickling1964,Ivany1965a,Mellen1956,Akulichev1968,Kedrinskii1972,Ebeling1978,Vokurka1986} published in the years and decades following the report of \citet{Kirkwood1942}. Many of these formidable works are difficult to obtain and are, thus, often not credited correctly or even overlooked in the modern literature. Further adding to the difficulties associated with gaining a comprehensive picture of the Kirkwood-Bethe hypothesis and models, the original report of Kirkwood and Bethe \citep{Kirkwood1942} does not appear to be indexed in scientific databases, meaning that a list of subsequent publications referencing this report is not readily available, and no review of the scientific literature related to the Kirkwood-Bethe hypothesis has been published.

The aim of this review is to gather, for the first time, the theoretical foundation of the Kirkwood-Bethe hypothesis, as well as models derived from it that are still relevant today, in a single document. Moreover, the description of the flow field arising from the Kirkwood-Bethe hypothesis is further generalized and the analysis of the limitations and assumptions underpinning the Kirkwood-Bethe hypothesis is extended. 

As a first step, Section \ref{sec:history} provides a short historical overview of major developments and applications related to the Kirkwood-Bethe hypothesis. Subsequently, the Kirkwood-Bethe hypothesis is derived from the conservation laws governing mass, momentum and energy in Section \ref{sec:kb}, paying particular attention to the underpinning assumptions. The application of different equations of state describing the properties of the liquid is presented in Section \ref{sec:eos}, the equation of motion of the gas-liquid interface of a pressure-driven gas bubble or cavity is presented in Section \ref{sec:bubble}, and models to describe the flow generated by oscillating or collapsing bubbles and cavities are derived in Section \ref{sec:liquid}. The model hierarchy with respect to the Mach number following from the Kirkwood-Bethe hypothesis is briefly reviewed in Section \ref{sec:hierarchy} and numerical methods to solve KBH models for the bubble dynamics, flow field and shock waves are discussed in Section \ref{sec:numerics}. In Section \ref{sec:results}, the assumptions and capabilities of the Kirkwood-Bethe hypothesis are further scrutinized and new results are presented. Conclusions are drawn and open questions are discussed in Section \ref{sec:conclusions}.\vspace{-1em}

\section{Retrospective}
\label{sec:history}

The Kirkwood-Bethe hypothesis was proposed as part of a technical report \citep{Kirkwood1942} commissioned by the research office of the United States federal government on predicting the pressure wave produced by an underwater explosion, completed in 1942 by John G.~Kirkwood and Hans A.~Bethe. This report focused on obtaining an approximation for the pressure amplitude at some distance from an underwater explosion and, more precisely, the decay constant describing an approximately exponential decay of the peak pressure. To this end, Kirkwood and Bethe focused on the propagation of the shock front, the conditions in the liquid behind the shock front, the rapidly expanding gas bubble of burnt detonation products, and the rarefaction wave forming inside the gas bubble. The work of Kirkwood and Bethe was first brought to the attention of a wider audience through the textbook of \citet{Cole1948}. 

\subsection{Theoretical foundations}

Shortly after its publication, \citet{Rice1943,Rice1944} extended the work of \citet{Kirkwood1942} to an explosive charge in the shape of an infinite cylinder by considering cylindrical instead of spherical symmetry. Although they found that the Kirkwood-Bethe hypothesis is less accurate in cylindrical symmetry than it is in spherical symmetry, it provided a first glimpse into the dynamics of underwater explosions that are not spherically symmetric. The work of \citet{Rice1943,Rice1944} later became the foundation of the studies of  \citet{Naugolnykh1971} on the hydrodynamics of electrical discharges in liquids and of \citet{Kedrinskii1972} on cylindrically symmetric underwater explosions. 

In 1952, \citet{Gilmore1952} studied the Kirkwood-Bethe hypothesis further, focusing, however, on cavitation bubble dynamics rather than on underwater explosions. Aside from clarifying some assumptions and exploring the theoretical limits of the Kirkwood-Bethe assumption, Gilmore presented a second-order differential equation of the radial motion of the gas-liquid interface of a spherical gas bubble in a compressible fluid as a result of ambient pressure differences, now known as the \textit{Gilmore model}, which has since become a staple of fundamental research on cavitation. As it contains important terms that are second order with respect to the Mach number of the gas-liquid interface motion, the Gilmore model presents an extension to the models of \citet{Herring1941} and \citet{Trilling1952} that are bound to first-order acoustics. In addition, \citet{Gilmore1952} also derived models for the flow velocity and pressure induced by a bubble or cavity under different assumptions. Among these models is an ordinary differential equation (ODE) for the flow velocity with respect to the radial coordinate along outgoing characteristics based on the Kirkwood-Bethe hypothesis,
which allows to reconstruct the flow and acoustics resulting from a bubble collapse or oscillation. Generalizing the Gilmore model for spherical bubbles, \citet{Kedrinskii1980} proposed a model of the radial dynamics of bubbles in planar and cylindrical symmetry.

\citet{Hickling1963} published an institutional report on a study of the collapse and rebound of an adiabatically compressed gas cavity in 1963, using the Kirkwood-Bethe hypothesis in conjunction with the method of characteristics. A shortened version of their report was published as a paper \citep{Hickling1964} the following year. To describe the flow of the liquid surrounding the gas cavity, Hickling and Plesset presented ODEs for both the flow velocity and pressure with respect to time along outgoing characteristics based on the Kirkwood-Bethe hypothesis, although they did not present a detailed derivation of these ODEs. Under the assumptions associated with the Kirkwood-Bethe hypothesis, the ODE for the flow velocity with respect to time ($\text{d}u/\text{d}t$) proposed by \citet{Hickling1963} is equivalent to the ODE with respect to the radial coordinate ($\text{d}u/\text{d}r$) of \citet{Gilmore1952}.

\citet{Ivany1965} revisited the theory on cavitation bubble dynamics and acoustic emissions in a PhD thesis on the collapse of gas cavities, completed in 1965. In this thesis, which was in parts published in a subsequent paper \citep{Ivany1965a}, Ivany presented a detailed derivation of an ODE for the flow velocity in the liquid starting from the momentum and continuity equations by using the Kirkwood-Bethe hypothesis and, in fact, arrived at the same ODE as \citet{Hickling1963} two years prior. However, while the model presented by Hickling and Plesset required to also solve an ODE for the pressure, Ivany employed an explicit expression for the pressure along outgoing characteristics obtained directly from the Kirkwood-Bethe hypothesis using the \citet{Tait1888} equation of state (EoS), as previously suggested by \citet{Gilmore1952}. In the appendix of the thesis, \citet{Ivany1965} carefully explains how the bubble dynamics and the flow field are solved and, remarkably, provides listings of the developed computer codes, written in the \textit{Michigan Algorithm Decoder} programming language, making it perhaps the first open-source software for cavitation research. Shortly after, in 1966, \citet{Lilliston1966} provided a computer code written in \textit{Fortran IV} for bubble dynamics and the resulting flow field based on the work of \citet{Gilmore1952} in the appendix of a report on the collapse of a gas-filled cavity.

\citet{Akulichev1968} proposed an explicit approximation for the pressure emitted by oscillating bubbles using Riemann invariants and the Tait EoS on the basis of the Kirkwood-Bethe hypothesis, which does not require explicit knowledge of the generated flow velocity. Moreover, while previous studies \citep{Gilmore1952, Hickling1963, Ivany1965} had deliberately avoided the multivalued solutions arising at shock fronts, \citet{Akulichev1968} used the \textit{rule of equal areas} \citep{Landau1959} to treat such multivalued solutions. As a result, Akulichev and co-workers were able to demonstrate that when the emitted pressure wave forms a shock front, the pressure may be severely overpredicted if the additional dissipation arising at the shock front is not taken into account. 

All the research reported in the publications discussed above employ the (modified) Tait equation of state \citep{Tait1888} to describe the liquid properties, starting with the original work of \citet{Kirkwood1942}. The main shortcoming of the Tait EoS is that it drastically overpredicts the temperature due to isentropic compression \citep{LeMetayer2016,Radulescu2020,Denner2021}. However, the Kirkwood-Bethe hypothesis is not limited to a specific EoS for the liquid. \citet{Flynn1957,Flynn1978} used a tabulated EoS based on the work of \citet{Rice1957} to define the properties of the liquid. Unfortunately, however, a copy of one of Flynn's reports \citep{Flynn1957} on this matter could not be located (the reference here relies on the account of \citet{Ivany1965}), while the other report \citep{Flynn1978} does not present the methodology in detail. Recently, \citet{Denner2021} extended the Gilmore model using the Noble-Abel stiffened-gas (NASG) EoS and, subsequently, \citet{Denner2023} used the NASG EoS successfully to model the flow and acoustics induced by cavitation bubbles in conjunction with the models of \citet{Gilmore1952}, \citet{Hickling1963}, and \citet{Ivany1965}.

These milestones form the foundation of KBH models today. Concurrent and subsequent research efforts have primarily focused on applying these models to various engineering problems, as well as to elucidate the physics of pressure-driven bubble dynamics and acoustics.

\subsection{Applications}

The application in the context of which the Kirkwood-Bethe hypothesis was originally conceived are underwater explosions, whereby the expansion of burnt gaseous detonation products and the emitted shock wave in the water are of primary interest. Early work on this subject, starting with the study of \citet{Kirkwood1942}, focused on predicting the evolution and decay of the peak pressure  of the emitted shock wave in spherical \citep{Kirkwood1942,Arons1954,McGrath1966,Poche1972,Rogers1977,Best1991,Shan2021,Zhang2021c} and cylindrical symmetry \citep{Rice1943,Rice1944,Kedrinskii1972,Likhterov2000}. To this end, the main aim was to find an accurate definition for the decay constant $\theta$ of the approximately exponential decay, $e^{-t/\theta}$, of the peak pressure. In the 21st century, the focus has shifted away from approximating the peak pressure in favor of using KBH models to compute and study the full range of dynamics arising in underwater explosions, including the associated bubble dynamics \citep{Kwak2012,Wang2021c,Zhang2021f,Li2023b}, pressure and shock waves \citep{Kwak2012,Sheng2023}, and fluid-structure interaction \citep{Mavaleix-Marchessoux2020}. These models have repeatedly been found to be of very high predictive quality \citep{Li2023b,Zhang2021f,Wang2021c,Sheng2023}. The research into underwater explosions is closely related to subsequent applications of the Kirkwood-Bethe hypothesis and, especially, the Gilmore model to predict the bubble dynamics of seismic air guns for geophysical exploration and the shock testing of defense vessels \citep{Ziolkowski1970,Ziolkowski1998,deGraaf2014,Bing-Shou2021}. For instance, \citet{Landro1992} proposed a methodology based on the Kirkwood-Bethe hypothesis to determine the pressure transients produced by an array of air guns, the results of which \citet{Laws1998} found to be in very good agreement with measurements in different experimental scenarios. Similar methods to predict the pressure field produced by air guns were proposed by \citet{Ziolkowski1970, Ziolkowski1998}. KBH models were also leveraged to study the influence of rough seas \citep{Sertlek2019} and shallow water \citep{Wehner2020} on the acoustic signatures of seismic air guns.

Other applications in which a strong local energy deposition produces a gaseous bubble that emits pressure waves, and subsequently collapses and rebounces, are cavitation bubbles induced by an electrical discharge or a focused laser beam. Especially laser-induced cavitation has established itself as a widely-used tool to study cavitation bubble dynamics \citep{Akhatov2001,Vogel2008,Supponen2016,Gonzalez-Avila2021,Bokman2023,Reuter2023,Zhao2023b,Mur2023,Wang2023a,Mnich2024,Fu2024} under controllable, precise, and reproducible conditions, both experimentally and computationally. \citet{Mellen1956} was an early adopter of the Kirkwood-Bethe hypothesis in the 1950s, in the form of the Gilmore model, to provide a theoretical basis for an experimental study of the evolution of a spherical cavity produced by an electrical discharge, reporting an encouraging agreement between experimental measurements and computational results. With respect to laser-induced cavitation, KBH models have been used, aside from studying the resulting bubble dynamics, to understand and quantify the formation and attenuation of shock waves  \citep{Vogel1996,Byun2004,Oh2018,Geng2021,Lai2022,Liang2022,Agrez2023,Wen2023a,Yang2023b} and the energy partitioning \citep{Liang2022,Wen2023a}. The results produced by these models have been shown to be in excellent agreement with experimental measurements \citep{Oh2018,Geng2021,Vassholz2021,Lai2022,Wen2023a}.

A major driver of research into pressure-driven bubble dynamics over the past decades have been diagnostic and therapeutic medical applications. In 1984, \citet{Vacher1984} used the Gilmore model to show that ultrasound-driven bubbles emit significant energy at the second harmonic of the driving frequency, now a widely used indicator to detect and classify cavitation in medical ultrasound treatments \citep{Versluis2020}. Since then, KBH models have been used to study tissue ablation \citep{Chavrier2000,BerlindaLaw2019,Wang2019}, histotripsy \citep{Pahk2015,Pahk2018,Pahk2019} and lithotripsy \citep{Sokolov2002} treatments, laser surgery \citep{Byun2004,Yang2023b}, and ultrasound-driven encapsulated microbubbles, including their dynamic behavior \citep{Ayme1989,Ayme-Bellegarda1990,Kreider2011,BerlindaLaw2019,Guemmer2021,Li2022a}, bubble-bubble interactions \citep{Qin2023a} and the associated acoustic emissions \citep{Kreider2011,Guemmer2021}. To this end, \citet{Guemmer2021} showed a favorable comparison of the frequency spectrum predicted by the Gilmore model for lipid-coated microbubbles with the experimental measurements of \citet{Song2019}. To account for the viscoelastic rheology of tissue, \citet{Zilonova2018,Zilonova2019} extended the Gilmore model to include the linear elastic solid model, also known as the Zener model, resulting in a markedly different bubble behavior \citep{Zilonova2018,Zilonova2019,Li2022a}.

The Kirkwood-Bethe hypothesis has also been utilized to study other applications of pressure-driven bubble dynamics, including food processing \citep{Bredihin2021}, the spall formation in aluminium \citep{Glam2014}, the treatment of materials \citep{Sonde2018,Zhang2024}, material synthesis \citep{Barcikowski2019,Huang2021,Huang2023}, the preferred nucleation of crystals during solidification \citep{Rakita2017}, surface cleaning \citep{Minsier2008}, as well as the dynamics of a collapsing cylindrical cavity driven by an annular piston \citep{Kedrinskii2022}. Studies on sonoluminescence, whereby ultrasound-driven bubbles emit a light pulse upon collapse \citep{Brenner2002}, have also directly benefited from KBH models, for instance to estimate the conditions insight the bubble during collapse \citep{Gimenez1982,Nazari-Mahroo2018,Hoeppe2024} and to correlate acoustic measurements to bubble dynamics \citep{Lee1997,Holzfuss1998,Holzfuss2010}. Aside from specific applications, KBH models have been used extensively to study and elucidate fundamental pressure-driven bubble dynamics and acoustics. Some examples of studies using the Gilmore model to study pressure-driven bubble dynamics include studies on the natural bubble frequency \citep{Shima1970}, self-similar bubble behavior \citep{Hunter1960,Akulichev1971}, period doubling of ultrasound-driven bubbles \citep{Akulichev1971}, the excitation of bubbles with multiple frequencies \citep{Moholkar2000}, as well as the influence of the liquid compressibility \citep{Fuster2011}, heat transfer \citep{Mahdi2010}, the vapor content \citep{Preso2024}, and a narrow confinement \citep{Zhang2023d} on the bubble dynamics. Moreover, the Kirkwood-Bethe hypothesis has been used to gain quantitative insight into the acoustic emissions associated with pressure-driven bubble dynamics, such as the acoustic noise spectrum generated by pressure-driven bubbles close to a free surface \citep{Likhterov1998,Berman2001}, the acoustic translation of bubbles \citep{Kim2023}, and the pressure and shock waves emitted during a strong bubble collapse \citep{Holzfuss2010,Liang2022,Denner2023,Wen2023a,Lai2022}. \vspace{-1em}

\section{The Kirkwood-Bethe hypothesis}
\label{sec:kb}

The Kirkwood-Bethe hypothesis can be derived using basic fluid dynamics, and the arising definitions and expressions then allow us to close the system of governing equations. In the following, we start in Section \ref{sec:kb_conservationlaws} by bringing the conservation laws governing a fluid flow into a convenient form. Subsequently, the Kirkwood-Bethe hypothesis is derived in Section \ref{sec:kb_kb} and its major limitations are discussed in Section \ref{sec:kb_limitations}.\vspace{-0.7em}

\subsection{Governing conservation laws}
\label{sec:kb_conservationlaws}

The conservation of mass, momentum and energy governing a fluid flow are given as
\begin{eqnarray}
    \frac{\text{D}\rho}{\text{D} t} + \rho \nabla \cdot \mathbf{u} &=& 0\\
     \frac{\text{D} \mathbf{u}}{\text{D} t}  + \frac{1}{\rho} \nabla p &=& \frac{1}{\rho} \, \nabla \cdot \boldsymbol{\tau} \\
     \frac{\text{D} h}{\text{D} t} - \frac{1}{\rho} \frac{\text{D}p}{\text{D}t} &=&   \frac{1}{\rho} \left(\boldsymbol{\tau} : \nabla \mathbf{u} - \nabla \cdot \dot{q} \right),
\end{eqnarray}
where $\rho$ is the fluid density, $\mathbf{u}$ is the flow velocity, $p$ denotes the pressure, $h$ is the specific enthalpy, $\boldsymbol{\tau}$ represents viscous stresses, $\dot{q}$ is the heat flux due to thermal conduction, and
\begin{equation}
    \frac{\text{D}}{\text{D} t} = \frac{\partial}{\partial t} + \left(\mathbf{u} \cdot \nabla \right) 
\end{equation}
denotes the material derivative operator.
The thermodynamic relation $\text{d}h = T \, \text{d}s + \text{d}p/\rho$ allows us to reformulate the energy equation in terms of the specific entropy $s$, such that
\begin{equation}
    \frac{\text{D} s}{\text{D} t} = \frac{1}{\rho T} \left(\boldsymbol{\tau} : \nabla \mathbf{u} - \nabla \cdot \dot{q} \right),
\end{equation}
where $T$ is the temperature. This system of equations is closed with a suitable equation of state (EoS) that defines a relationship between the thermodynamic quantities. Since the subsequent derivation is agnostic to the applied EoS, we proceed without assuming a specific EoS and revisit commonly used EoS in Section \ref{sec:eos}.

Assuming that the considered flow is ideal, without viscous dissipation ($\boldsymbol{\tau} = 0$) and heat conduction ($\dot q = 0$), we obtain the Euler equations:
\begin{eqnarray}
    \frac{\text{D}\rho}{\text{D} t} + \rho \nabla \cdot \mathbf{u} &=& 0 \label{eq:continuity_3d_simple}\\
     \frac{\text{D} \mathbf{u}}{\text{D} t}  + \frac{1}{\rho} \nabla p &=& 0 \label{eq:momentum_3d_simple}\\
     \frac{\text{D} s}{\text{D} t} &=& 0. \label{eq:entropy_3d_simple}
\end{eqnarray} 
This set of governing equations, with which \citet{Kirkwood1942} commenced their work, assumes that changes to the state of the fluid of a sufficiently smooth flow are adiabatic and reversible. Consequently, the flow is \textit{isentropic}, as imposed by Eq.~\eqref{eq:entropy_3d_simple}. Especially for water, this turns out to be a good assumption even for pressure waves with a large amplitude, because the resulting change in entropy is small. The previously suggested upper limit for the validity of this approximation is $p = \mathcal{O}(\text{GPa})$.\citep{Kirkwood1942,Hickling1964} Assuming that the flow is isentropic does, however, not imply that all fluid elements are on the same isentrope, because the entropy may change at a shock front, which is further discussed in Section \ref{sec:numerics_shocks}.

The differential relation of the specific enthalpy for an isentropic fluid is 
\begin{equation}
    \mathrm{d}h = \frac{\mathrm{d}p}{\rho} = c^2 \, \frac{\mathrm{d}\rho}{\rho}, \label{eq:dh_dp_drho}
\end{equation}
with the speed of sound $c$ defined as
\begin{equation}
    c^2 = \left(\frac{\partial p}{\partial \rho}\right)_{s},
\end{equation}
where subscript $s$ implies a constant entropy. In addition, to account for a time-varying pressure in the far field, we introduce the ambient enthalpy $h_\infty$ and redefine, in a slight abuse of notation, the specific enthalpy difference as $h \rightarrow h - h_\infty$. The ambient enthalpy $h_\infty$ is spatially invariant ($\nabla h_\infty =0$), such that
\begin{equation}
    \frac{\text{D}h_\infty}{\text{D}t} =  \frac{\partial h_\infty}{\partial t}. \label{eq:dhinf_dt}
\end{equation}
Inserting Eq.~\eqref{eq:dh_dp_drho} in the form $\text{d}\rho = \rho \, \text{d}h/c^2$ alongside the ambient enthalpy $h_\infty$ into the continuity equation, Eq.~\eqref{eq:continuity_3d_simple}, yields
\begin{equation}
    \frac{1}{c^2} \, \frac{\text{D}h}{\text{D}t} + \nabla \cdot \mathbf{u} = \frac{1}{c^2} \, \frac{\partial h_\infty}{\partial t}. \label{eq:continuity_3d_h}
\end{equation} 
For convenience, we also separate the kinetic energy and the rotation of the flow by reformulating the convective velocity derivative as
\begin{equation}
    \left( \mathbf{u} \cdot \nabla \right) \mathbf{u} =  \nabla \left(\frac{\mathbf{u}^2}{2}\right) - \mathbf{u} \times (\nabla \times \mathbf{u}), \label{eq:convderivative_untangled}
\end{equation}
with which the momentum equation, Eq.~\eqref{eq:momentum_3d_simple}, becomes
\begin{eqnarray}
    \frac{\partial \mathbf{u}}{\partial t} + \nabla \left(\frac{\mathbf{u}^2}{2}\right) - \mathbf{u} \times (\nabla \times \mathbf{u}) + \nabla h &=& 0. \label{eq:momentum_3d_h}
\end{eqnarray}
Note that, since the density and pressure have been replaced by the specific enthalpy and the speed of sound on the basis of the flow being isentropic, the entropy equation, Eq.~\eqref{eq:entropy_3d_simple}, is now obsolete.

\subsection{Hypothesis}
\label{sec:kb_kb}

\begin{figure*}
    \includegraphics[scale=1]{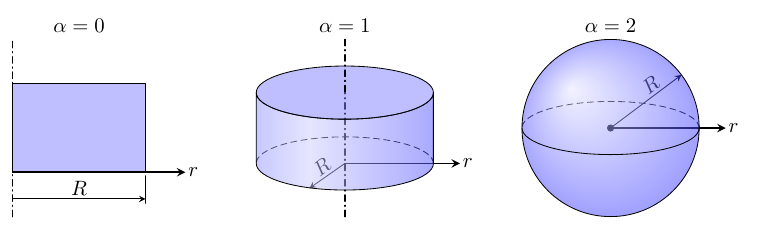}
    \caption{{Schematic illustration of the considered (quasi) one-dimensional domains: one-dimensional planar ($\alpha = 0$), cylindrical symmetry ($\alpha=1$), and spherical symmetry ($\alpha=2$). The one-dimensional spatial coordinate is denoted with $r$ and the width or radius of the gas cavity or bubble is denoted with $R$.}}
    \label{fig:domains_schematic}
\end{figure*}

The Kirkwood-Bethe hypothesis is typically derived and applied under the assumption of spherical symmetry. However, it can be derived more generally for one-dimensional problems, e.g.~exploiting cylindrical symmetry \citep{Rice1943}, to explore the influence of the shape of a bubble on its dynamics and the induced flow field. Here, we consider an arbitrary (quasi) one-dimensional domain, {illustrated in Figure \ref{fig:domains_schematic},} characterized by the dimensionality coefficient $\alpha$ that defines the effective dimensions of the system: $\alpha = 0$ describing a planar case, $\alpha=1$ describing cylindrical symmetry, and $\alpha=2$ describing the conventionally assumed spherical symmetry.

Considering a (quasi) one-dimensional flow by exploiting the symmetry of the system, we know that the flow is \textit{irrotational}, with $\nabla \times \mathbf{u}=0$. The continuity equation, Eq.~\eqref{eq:continuity_3d_h}, and the momentum equation, Eq.~\eqref{eq:momentum_3d_h}, along the one-dimensional spatial coordinate $r$ are, therefore, given as
\begin{eqnarray}
    \frac{1}{c^2} \left(\frac{\partial h}{\partial t} + u \, \frac{\partial h}{\partial r} \right) + \frac{1}{r^\alpha} \, \frac{\partial}{\partial r} \left(r^\alpha u \right) &=&  \frac{1}{c^2} \, \frac{\partial h_\infty}{\partial t} \label{eq:continuity_h} \\
    \frac{\partial u}{\partial t} + \frac{\partial}{\partial r} \left(\frac{u^2}{2}\right) + \frac{\partial h}{\partial r} &=& 0.
    \label{eq:momentum_h}
\end{eqnarray}
Since the flow is irrotational, the velocity may be expressed in terms of the potential $\psi$, 
\begin{equation}
    u = -\frac{\partial \psi}{\partial r},  \label{eq:u_potential}
\end{equation}
with which the continuity and momentum equations can be written as
\begin{eqnarray}
    \frac{1}{c^2} \left(\frac{\partial h}{\partial t} - \frac{\partial h_\infty}{\partial t} + u \, \frac{\partial h}{\partial r} \right) -  \frac{1}{r^\alpha} \frac{\partial}{\partial r} \left(r^\alpha \frac{\partial \psi}{\partial r}\right) \!\!\! &=& \!\! 0 \label{eq:continuity_potential}\\
   - \frac{\partial}{\partial t} \frac{\partial \psi}{\partial r} + \frac{\partial}{\partial r} \left(\frac{u^2}{2}\right) + \frac{\partial h}{\partial r} \!\!\! &=& \!\! 0. \label{eq:momentum_potential}
\end{eqnarray}
Integrating Eq.~\eqref{eq:momentum_potential} from $r$ to $\infty$, assuming the potential $\psi$ vanishes at infinity, yields the transient Bernoulli equation,
\begin{equation}
    -\frac{\partial \psi}{\partial t} + \frac{u^2}{2} + h- h_\infty= 0,
    \label{eq:momentum_int}
\end{equation}
based on which \citet{Kirkwood1942} defined the specific kinetic enthalpy as
\begin{equation}
    \Omega = \frac{\partial \psi}{\partial t} =  h-h_\infty + \frac{u^2}{2}.  \label{eq:omega}
\end{equation}

Rearranging Eq.~\eqref{eq:omega} for the enthalpy difference $h-h_\infty$ and inserting the resulting expression into the continuity equation, Eq.~\eqref{eq:continuity_potential}, yields
\begin{multline}
    \frac{1}{c^2} \left(\frac{\partial^2 \psi}{\partial t^2} + u \frac{\partial }{\partial r} \frac{\partial \psi}{\partial t} - \frac{1}{2} \frac{\partial u^2}{\partial t} -  \frac{1}{2} u \frac{\partial u^2}{\partial r}\right) \\ -  \frac{1}{r^\alpha} \frac{\partial}{\partial r} \left(r^\alpha \frac{\partial \psi}{\partial r}\right) = 0. \label{eq:continuity_potential_wave}
\end{multline}
Noting that the last term on the left-hand side is the Laplacian of the velocity potential, 
\begin{equation}
    \nabla^2 \psi = \frac{1}{r^\alpha} \frac{\partial}{\partial r} \left(r^\alpha \frac{\partial \psi}{\partial r}\right), \label{eq:laplacian}
\end{equation}
Eq.~\eqref{eq:continuity_potential_wave} can be rewritten as a wave equation with a nonlinear correction of order $\mathcal{O}(u^2)$, \citep{Kirkwood1942,Keller1956}
\begin{equation}
     \frac{1}{c^2} \, \frac{\partial^2 \psi}{\partial t^2} - \nabla^2 \psi = \frac{1}{c^2} \left(\frac{\partial u^2}{\partial t} + \frac{1}{2} u \frac{\partial u^2}{\partial r} \right),
     \label{eq:kb_wave}
\end{equation}
suggesting that the velocity potential $\psi$ propagates like a wave. As discussed in detail by \citet{Prosperetti1984a}, the finite propagation speed is unimportant in the near field, where the flow behaves similar to an incompressible fluid ($c \rightarrow \infty$), with $\nabla^2 \psi = 0$, while the perturbations are small in the far field, such that terms of order $\mathcal{O}(u^2)$ become negligible \citep{Akulichev1968}, and Eq.~\eqref{eq:kb_wave} reduces to the classic form of the wave equation,
\begin{equation}
    \frac{1}{c^2} \, \frac{\partial^2 \psi}{\partial t^2} - \nabla^2 \psi = 0.
     \label{eq:wave_acoustic}
\end{equation}
Comparing the magnitude of the right-hand side of Eq.~\eqref{eq:kb_wave} to the magnitude of the second-order time derivative of the velocity potential on the left-hand side may be used to quantify the local error of using a wave equation to propagate acoustic information in this case \citep{Mavaleix-Marchessoux2020}.  

The work of \citet{Kirkwood1942} rests on a reformulated velocity potential in spherical symmetry, defined as $\phi = r \psi$. With this new potential, the wave equation, Eq.~\eqref{eq:kb_wave}, becomes
\begin{equation}
    \frac{1}{c^2} \, \frac{\partial^2 \phi}{\partial t^2} - \frac{\partial^2 \phi}{\partial r^2} = \frac{r}{c^2} \left(\frac{\partial u^2}{\partial t} + \frac{1}{2} u \frac{\partial u^2}{\partial r} \right)
\end{equation}
and the transient Bernoulli equation, Eq.~\eqref{eq:momentum_int}, yields
\begin{equation}
    \frac{\partial \phi}{\partial t} = r \, \frac{\partial \psi}{\partial t} = r \left(  h- h_\infty + \frac{u^2}{2} \right),
\end{equation}
which is, in fact, the quantity defined in Eq.~\eqref{eq:kb}. Adopting this principle for the general one-dimensional domain considered here, with $\alpha$ defining the symmetry of the problem, the reformulated velocity potential is defined as $\phi = r^{\alpha/2}\psi$, originally proposed by \citet{Rice1943}. The wave equation then reads as
\begin{equation}
    \frac{1}{c^2} \, \frac{\partial^2 \phi}{\partial t^2} - \frac{\partial^2 \phi}{\partial r^2} = \frac{r^{\alpha/2}}{c^2}  \left(\frac{\partial u^2}{\partial t} + \frac{1}{2} u \frac{\partial u^2}{\partial r} \right) \label{eq:kb_wave_general}
\end{equation}
and the transient Bernoulli equation yields
\begin{equation}
    \frac{\partial \phi}{\partial t} = r^{\alpha/2} \, \frac{\partial \psi}{\partial t} = r^{\alpha/2} \left(h- h_\infty + \frac{u^2}{2}\right). \label{eq:kb_general}
\end{equation}
The validity of this definition of the velocity potential is further discussed together with the limitations presented by the dimensionality coefficient $\alpha$ in Section \ref{sec:kb_limitations}.

In principle, the method of characteristics may be applied to solve the wave equation governing the velocity potential and any of its derivatives, even though an analytical solution is not readily obtainable due to the contributions of $\mathcal{O}(u^2)$.\citep{Akulichev1968} 
The \textit{Kirkwood-Bethe hypothesis} stipulates that the quantity $g={\partial \phi}/{\partial t}$, as given by Eq.~\eqref{eq:kb_general}, is constant along an outgoing characteristic, satisfying \citep{Kirkwood1942}
\begin{equation}
  \left. \frac{\mathrm{d}g}{\mathrm{d}t} \right|_\mathrm{c} = 0,
  \label{eq:kb_char}
\end{equation}
where the derivative operator along an outgoing characteristic is defined as
\begin{equation}
    \left. \frac{\mathrm{d}}{\mathrm{d}t}\right|_\mathrm{c} = \frac{\partial }{\partial t} + \left. \frac{\mathrm{d}r}{\mathrm{d}t}\right|_\mathrm{c}  \frac{\partial }{\partial r}. \label{eq:derivative_char}
\end{equation}
Observing that small disturbances to the velocity potential of the fluid propagate like a wave according to Eq.~\eqref{eq:kb_wave}, i.e.~with speed $c$ relative to the flow, \citet{Kirkwood1942} proposed to approximate the propagation speed  along an outgoing characteristic by
\begin{equation}
    \left. \frac{\mathrm{d}r}{\mathrm{d}t}\right|_\mathrm{c} = c+u. \label{eq:drdt_char}
\end{equation}

\subsection{Limitations}
\label{sec:kb_limitations}

The assumption of a constant value of $g = r^{\alpha/2} (h-h_\infty + u^2/2)$ along outgoing characteristics can be justified by the conservation of specific energy when dissipative effects are neglected, as it follows from the transient Bernoulli equation. However, the assumptions that the acoustic potential is $\phi=r^{\alpha/2} \psi$ and propagates with speed $c+u$ may entail errors.

A plane finite-amplitude acoustic wave moving in one direction (\textit{cf.}~simple wave) propagates with speed $c+u$, \citep{Rayleigh1910} and $c \pm u$ are the eigenvalues of the Euler equations for plane waves,\citep{LeVeque2008} meaning that plane waves travel with the local speed of sound $c$ relative to the flow \citep{Landau1959}. Moreover, Eq.~\eqref{eq:drdt_char} reduces to the asymptotic speed $c$ in the far field, since $u\rightarrow0$ for $r\rightarrow\infty$, and is consistent with the Chapman-Jouguet condition \citep{Chapman1899,Jouguet1905} for detonation waves. However, the propagation speed $c+u$ is not exact. This can be seen by considering the Riemann function $\varsigma$ for an isentropic flow \citep{Riemann1860},
\begin{equation}
    \varsigma = \int_{p_\infty}^p \frac{\text{d}p}{\rho c} = \int_{h_\infty}^h \frac{\text{d}h}{c} . \label{eq:riemann_function}
\end{equation}
Introducing $\varsigma$ into the continuity and momentum equations, Eqs.~\eqref{eq:continuity_h} and \eqref{eq:momentum_h}, yields 
\begin{eqnarray}
    \frac{\partial }{\partial t} (\varsigma + u) + (c + u) \frac{\partial}{\partial r} (\varsigma + u) &=& -\frac{\alpha u c}{r} \label{eq:riemann_outgoing}\\
    \frac{\partial }{\partial t} (\varsigma - u) + (c - u) \frac{\partial}{\partial r} (\varsigma - u) &=& -\frac{\alpha u c}{r},
\end{eqnarray}
with the quantities $\varsigma + u$ and $\varsigma - u$ being so-called \textit{Riemann invariants} \citep{Riemann1860}.
For a plane wave ($\alpha=0$), the Riemann invariant $\varsigma + u$  propagates with constant value along the right-going characteristic with speed $c + u$, whereas $\varsigma - u$ propagates with constant value along the left-going characteristic with speed $c - u$. When the right-hand side, which accounts for the geometric divergence of, for instance, cylindrical ($\alpha=1$) or spherical ($\alpha=2$) domains, is not zero, the Riemann invariants $\varsigma + u$ and $\varsigma - u$ do not propagate with $c+u$ and $c-u$, respectively. Hence, quantities that are invariant along outgoing characteristics of a curved wave do not propagate with speed $c+u$, rendering Eq.~\eqref{eq:drdt_char} an approximation that becomes more accurate as a curved wave moves outwards or when the velocity is small with respect to the speed of sound \citep{Gilmore1952}. Assuming the (dominant) fluid flow is the result of the bubble motion and the associated acoustic emissions only, the velocity is $u \propto r^{-\alpha/2}$ in the far field and, hence, the right-hand side of Eq.~\eqref{eq:riemann_outgoing} decays rapidly with $r^{-n}$, where $n \geq 1+\alpha/2$.

Considering a plane acoustic wave, the acoustic particle velocity $u_1 = p_1/(\rho c)$ is in phase with the acoustic pressure $p_1$, whereby $p_1$ and $u_1$ follow from the expansions $p=p_0+p_1$ and $u=u_0 + u_1$, and where subscript $0$ denotes the reference or ambient properties of the flow field. However, $u_1$ and $p_1$ of a curved wave are not in phase \citep{Kinsler2000}. As a consequence, the specific acoustic impedance $z=p_1/u_1$ experienced by a curved wave is complex, given for a spherical harmonic wave as \citep{Kinsler2000,Randall2005}
\begin{equation}
    z = \rho c \, \frac{(k r)^2 + i kr}{1+(k r)^2} , \label{eq:impedance}
\end{equation}
which is dependent on the wavenumber $k$ and radial coordinate $r$, and where $i$ denotes the imaginary unit. For large $r$, the imaginary part (specific acoustic reactance) vanishes and the real part (specific acoustic resistance) approaches the specific acoustic impedance of a plane wave, $z=\rho c$, whereas the imaginary part dominates for small $r$. KBH models do not account for this complex acoustic impedance and the associated error is investigated in Section \ref{sec:results_impedance}. 

The dimensionality coefficient $\alpha$ defines the dimension and symmetry of the considered problem. Inserting $\phi=r^{\alpha/2} \psi$ in the corresponding wave equation, Eq.~\eqref{eq:kb_wave_general}, reads as
\begin{multline}
    \frac{1}{c^2} \, \frac{\partial^2 (r^{\alpha/2} \psi)}{\partial t^2} - \frac{\partial^2 (r^{\alpha/2} \psi)}{\partial r^2} = \frac{r^{\alpha/2}}{c^2}  \left(\frac{\partial u^2}{\partial t} + \frac{1}{2} u \frac{\partial u^2}{\partial r} \right) ,
\end{multline}
which, after expanding the left-hand side and dividing by $r^{\alpha/2}$,   becomes
\begin{multline}
    \frac{1}{c^2} \, \frac{\partial^2 \psi}{\partial t^2}  - \underbrace{\left(\frac{\partial^2 \psi}{\partial r^2} + \frac{\alpha}{r} \frac{\partial \psi}{\partial r}\right)}_{\nabla^2 \psi,\text{~Eq.~\eqref{eq:laplacian}}}  - \frac{\alpha (\alpha - 2)}{4r^2} \psi \\ = \frac{1}{c^2}  \left(\frac{\partial u^2}{\partial t} + \frac{1}{2} u \frac{\partial u^2}{\partial r} \right) . \label{eq:kb_wave_general_expanded}
\end{multline}
The last term on the left-hand side vanishes for plane ($\alpha=0$) and spherical ($\alpha=2$) waves, such that Eq.~\eqref{eq:kb_wave_general_expanded} is identical to Eq.~\eqref{eq:kb_wave}. This means that the potential $\psi$ and the shape of plane and spherical waves of small amplitude, where $u^2 \ll c^2$, remain unchanged as they propagate away from the emitter. However, the corresponding wave equation for cylindrical waves ($\alpha = 1$), 
\begin{equation}
    \frac{1}{c^2} \, \frac{\partial^2 \psi}{\partial t^2}  - \nabla^2 \psi + \frac{\psi}{4r^2} = \frac{1}{c^2}  \left(\frac{\partial u^2}{\partial t} + \frac{1}{2} u \frac{\partial u^2}{\partial r} \right) , \label{eq:kb_wave_general_expanded_cyl}
\end{equation}
contains a spurious term proportional to $1/r^2$ and provides an accurate solution only in the asymptotic limit $r\rightarrow \infty$. \citep{Rice1943,Kedrinskii2005} 
In fact, in the interval $0 \leq \alpha \leq 2$, the spurious term in Eq.~\eqref{eq:kb_wave_general_expanded} is largest for $\alpha=1$, and \citet{Rice1943,Rice1944} reported more accurate results for cylindrical underwater explosions with $\alpha=0.8$.

In summary, the assumptions underpinning the Kirkwood-Bethe hypothesis are, strictly speaking, valid only when either the wavelength is relatively small, such that the plane wave assumption is valid \citep{Blackstock1964}, or when the Mach number of the induced flow is small. The various modelling errors of the Kirkwood-Bethe hypothesis vanish in the asymptotic limits $r \rightarrow \infty$, representing the far field, or $u \rightarrow 0$, representing the linear acoustic regime. More generally, the Kirkwood-Bethe hypothesis can be regarded as an extrapolation of the linear acoustic theory \citep{Prosperetti1984a} and, as briefly revisited in Section \ref{sec:hierarchy}, is part of a consistent hierarchy of models \citep{Denner2023,Zhang2023b}. Nevertheless, as further discussed in Section \ref{sec:results}, the Kirkwood-Bethe hypothesis provides accurate predictions under the specific conditions associated with pressure-driven bubble dynamics, cavitation and underwater explosions.

\section{Liquid equation of state}
\label{sec:eos}

The Kirkwood-Bethe hypothesis does not assume a specific equation of state (EoS) for the liquid, even though it is almost always presented and used in conjunction with the modified Tait EoS \citep{Tait1888}, which \citet{Kirkwood1942} used in their original work. Similarly, the influential works of \citet{Cole1948} on underwater explosions and \citet{Gilmore1952} on cavitation also used the modified Tait EoS. The {reasons} for the popularity of this EoS are rather straightforward: it has a simple analytical form and it provides an accurate (isentropic) relationship between the pressure $p$, the specific enthalpy $h$, the density $\rho$, and the speed of sound $c$ of water, even when compared against modern standards, such as the IAPWS R6-95(2018) standard \citep{Wagner2002}.

\begin{figure*}
  \includegraphics[width=0.85\linewidth]{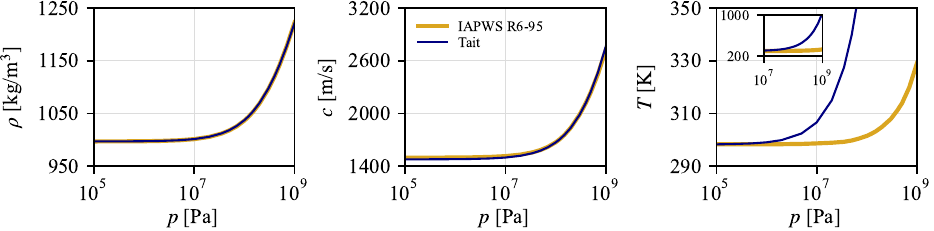}
  \caption{Density $\rho$, speed of sound $c$, and temperature $T$ of water obtained with the modified Tait EoS, using the model constants of \citet{Cole1948} ($n=7.15$ and $B=3047 \times 10^5 \, \text{Pa}$), compared against the IAPWS R6-95(2018) standard. The reference values are $p_0 = 10^5 \, \text{Pa}$, $\rho_0 = 997 \, \text{kg/m}^3$, and $T_0 = 298.3 \, \text{K}$.}
  \label{fig:tait}
\end{figure*}

While the original Tait EoS is formulated along isotherms \citep{Tait1888},
\begin{equation}
    - \left( \frac{\partial v}{\partial p} \right)_T = \frac{1}{n \left[ p+B(T)\right]},
\end{equation}
where $v=1/\rho$ is the specific volume, $n$ is the polytropic exponent, and $B$ is a pressure constant, the modified Tait EoS used widely in conjunction with the Kirkwood-Bethe hypothesis is formulated along isentropes \citep{Cole1948}
\begin{equation}
    - \frac{1}{v} \left( \frac{\partial v}{\partial p} \right)_s = \frac{1}{n \left[ p+B(s)\right]},
\end{equation}
which is, in essence, the isentropic (or, more generally, polytropic) form of the stiffened-gas EoS \citep{Harlow1971,LeMetayer2004}. The modified Tait EoS defines the fluid properties based on the pressure $p$ as
\begin{eqnarray}
    h &=& \frac{n}{n-1} \frac{p+B}{\rho}  \label{eq:h_Tait} \\
    \rho &=& \rho_0 \left(\frac{p+B}{p_0+B}\right)^{\frac{1}{n}} \label{eq:rho_Tait}\\
    c &=&\sqrt{n \, \frac{p+B}{\rho}},\label{eq:c_Tait}
\end{eqnarray}
where subscript $0$ defines a reference state. For water, commonly used values are $n \approx 7$ and $B \approx 3 \times 10^8 \, \text{Pa}$, or $B \approx \rho_0 c_0^2 / n$.\citep{Kirkwood1942} 
The reference density, pressure and speed of sound of water are typically assumed to be $\rho_0 \approx 1000 \, \text{kg/m}^3$, $p_0 \approx 10^5 \, \text{Pa}$, and $c_0 \approx 1500 \, \text{m/s}$, respectively.  Figure \ref{fig:tait} shows the density, speed of sound and temperature predicted by the Tait EoS in comparison to the IAPWS R6-95(2018) standard.

The polytropic exponent for an isentropic flow represents the ratio of specific heats, which is approximately unity for water and other liquids. Thus, the large polytropic exponent for water, $n \approx 7$, points to the main shortcoming of the Tait EoS: it severely overpredicts changes in temperature as a result of adiabatic compression or expansion \citep{Radulescu2020,Denner2021},
\begin{equation}
    T = T_0 \left(\frac{p+B}{p_0+B}\right)^{(n-1)/n}, \label{eq:T_EoS}
\end{equation}
as seen in Figure \ref{fig:tait}. In fact, during a violent bubble collapse, the Tait EoS may predict the temperature of the liquid in the vicinity of the gas-liquid interface to be higher than the temperature inside the gas bubble \citep{Denner2021}, which is unphysical on account of the considerably larger heat capacity of liquids compared to gases.

\begin{table*}
    \caption{Model constants of the NASG EoS for water proposed in recent years.}
    \label{tab:nasg}
    \begin{tabular}{l|ccc}
         & ~\citet{LeMetayer2016}~ & ~\citet{Chandran2019}~ & ~\citet{Denner2023}~\\
        \hline 
        $n$ & 1.19 & 1.19 & 1.11\\
        $B \ [\text{Pa}]$ & $7028 \times 10^5$ & $6218 \times 10^5$ & $6480 \times 10^5$ \\
        $b \ [\text{m}^3/\text{kg}]$ & $6.61 \times 10^{-4}$ & $6.72 \times 10^{-4}$ & $6.80 \times 10^{-4}$\\
        $\rho_0 \ [\text{kg/m}^3]$ & $957.7$ & $997.0$ & $997.0$ \\
        $p_0 \ [\text{Pa}]$ & $1.0453 \times 10^5$ & $10^5$ & $10^5$\\
        \hline
    \end{tabular}
\end{table*}

\begin{figure*}
    \includegraphics[width=0.85\linewidth]{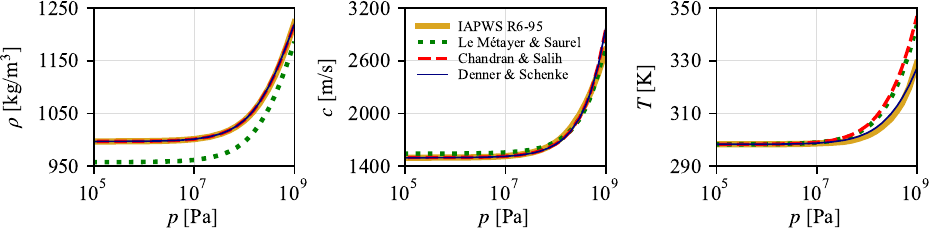}
    \caption{Density $\rho$, speed of sound $c$, and temperature $T$ of water obtained with the NASG EoS, using the model constants listed in Table \ref{tab:nasg}, compared against the IAPWS R6-95(2018) standard. The reference temperature is $T_0 = 298.3 \, \text{K}$.}
    \label{fig:nasg}
\end{figure*}

To rectify this shortcoming of the Tait EoS and, similarly, the stiffened-gas EoS, \citet{LeMetayer2016} proposed the Noble-Abel stiffened-gas (NASG) EoS, which had already been postulated in similar (isothermal) form by \citet{Tammann1912}. The NASG EoS extends the Tait and stiffened-gas EoS by introducing the co-volume $b$ that represents the finite volume occupied by the fluid molecules, rendering the NASG EoS a simplified version of the van-der-Waals EoS \citep{Kontogeorgis2019} with constant coefficients $B$ and $b$. For an isentropic flow the specific enthalpy, density and speed of sound are defined as 
\citep{Denner2021}
\begin{eqnarray}
    h &=& \frac{n}{n-1} \frac{p+B}{\rho} - \frac{n \, b}{n-1} \, (p+B) + b \, p \label{eq:h_NASG} \\
    \rho &=& \frac{K \, (p+B)^{\frac{1}{n}}}{1+b \, K \,  (p+B)^{\frac{1}{n}}} \label{eq:rho_NASG}\\
      c &=&\sqrt{n \, \frac{p+B}{\rho-b  \rho^2}},\label{eq:c_NASG}
\end{eqnarray}
where $K = \rho_0/[(p_0+B)^{{1/n}} \ (1-b \, \rho_0)]$ describes a constant reference state. Just like the Tait EoS, to which it reduces for $b=0$, the NASG EoS is unconditionally convex \citep{LeMetayer2016}, making it an equally robust alternative to the Tait EoS. The temperature as a result of adiabatic compression and expansion is also for the NASG EoS given by Eq.~\eqref{eq:T_EoS}. Different model constants for the NASG EoS of water have been proposed in recent years, as listed in Table \ref{tab:nasg} and shown in Figure \ref{fig:nasg}. In particular the model constants for isentropic water proposed by \citet{Denner2023} exhibit a very good agreement with the IAPWS R6-95(2018) standard. The NASG EoS was first used in conjunction with the Kirkwood-Bethe hypothesis in the Gilmore-NASG model for bubble dynamics by \citet{Denner2021} and, using the theory derived in Section \ref{sec:liquid} for spherical bubbles, to model the full flow field introduced by cavitation bubbles by \citet{Denner2023}.

Other equations of state, such as the van-der-Waals EoS or any tabulated EoS, may also be applied in conjunction with the Kirkwood-Bethe hypothesis. However, the simplicity and robustness of the Tait EoS and, more recently, the NASG EoS have prevailed so far. 

\section{Bubble dynamics}
\label{sec:bubble}

The propagation of the invariant $g$ along outgoing characteristics, as stipulated by the Kirkwood-Bethe hypothesis, can be readily employed to derive the equation of motion of the gas-liquid interface of a gas bubble or cavity.

Expanding Eq.~\eqref{eq:kb_char} under consideration of the time-dependent ambient enthalpy defined by Eq.~\eqref{eq:dhinf_dt} yields
\begin{multline}
    r \frac{\text{D}h}{\text{D} t} - r \frac{\text{D}h_\infty}{\text{D} t} + r u \frac{\text{D}u}{\text{D} t} + r c \left(\frac{\partial h}{\partial r} + u \frac{\partial u}{\partial r}\right) \\ + \frac{\alpha}{2} \left(h-h_\infty+\frac{u^2}{2}\right) \left. \frac{\mathrm{d}r}{\mathrm{d}t}\right|_\mathrm{c} = 0.
    \label{eq:kirkwoodbethe_char_2}
\end{multline}
Expressions for the spatial derivatives of the flow velocity and specific enthalpy follow by rearranging the continuity and momentum equations, Eqs.~\eqref{eq:continuity_h} and \eqref{eq:momentum_h}, as
\begin{eqnarray}
     \frac{\partial u}{\partial r} &=& -\frac{1}{c^2} \left(\frac{\text{D}h}{\text{D} t} - \frac{\text{D}h_\infty}{\text{D} t}  \right) - \frac{\alpha u}{r} \label{eq:dudr}\\
     \frac{\partial h}{\partial r}  &=& - \frac{\text{D}u}{\text{D} t}, \label{eq:dhdr}
\end{eqnarray}
respectively. Inserting Eqs.~\eqref{eq:drdt_char}, \eqref{eq:dudr}, and \eqref{eq:dhdr} into Eq.~\eqref{eq:kirkwoodbethe_char_2}, we obtain
\begin{multline}
    r \left(\frac{\text{D}h}{\text{D} t} - \frac{\text{D}h_\infty}{\text{D} t}\right) + r \frac{\text{D}u}{\text{D} t} (u-c) - \frac{ru}{c} \left(\frac{\text{D}h}{\text{D} t} - \frac{\text{D}h_\infty}{\text{D} t}  \right)  \\ - \alpha {c u^2} + \frac{\alpha}{2} \left(h-h_\infty+\frac{u^2}{2}\right) (c+u) = 0. \label{eq:kirkwoodbethe_char_3}
\end{multline}
Evaluating this equation at the gas-liquid interface ($r=R$) yields a second-order ODE describing the change in radius $R$ of a gas bubble or cavity, given as
\begin{multline}
    \left( 1 - \frac{\dot{R}}{c_{\mathrm{L}}} \right) R \ddot{R} + \frac{3}{4} \alpha \left( 1 - \frac{\dot{R}}{3 c_{\mathrm{L}}} \right) \dot{R}^2 \\ = \frac{\alpha}{2} \left( 1 + \frac{\dot{R}}{c_{\mathrm{L}}} \right) H + \left( 1- \frac{\dot{R}}{c_{\mathrm{L}}} \right) \frac{R \dot{H}}{c_{\mathrm{L}}}, \label{eq:bubbleODE}
\end{multline}  
with $H = h_{\mathrm{L}} - h_\infty$ and $\dot{H} = \dot{h}_{\mathrm{L}} - \dot{h}_\infty$, where $\dot{\lozenge} = {\text{D}\lozenge}/{\text{D} t}$ denotes the material derivative at the gas-liquid interface and the subscript $\mathrm{L}$ denotes quantities of the liquid at the gas-liquid interface. In this general form, Eq.~\eqref{eq:bubbleODE} was first presented by \citet{Kedrinskii1980}, while in spherical symmetry ($\alpha = 2$), Eq.~\eqref{eq:bubbleODE} is the well-known Gilmore equation \citep{Gilmore1952}
\begin{multline}
    \left( 1 - \frac{\dot{R}}{c_{\mathrm{L}}} \right) R \ddot{R} + \frac{3}{2} \left( 1 - \frac{\dot{R}}{3 c_{\mathrm{L}}} \right) \dot{R}^2 \\ = \left( 1 + \frac{\dot{R}}{c_{\mathrm{L}}} \right) H + \left( 1- \frac{\dot{R}}{c_{\mathrm{L}}} \right) \frac{R \dot{H}}{c_{\mathrm{L}}}. \label{eq:bubbleODE_gilmore}
\end{multline} 
Note that the assumption of a cylindrical or spherical bubble, as well as the assumption of a spatially invariant ambient enthalpy $h_\infty$, are only valid if $\lambda_\infty \gg R$, \citep{Neppiras1980, Tilmann1980} where $\lambda_\infty$ is the wavelength of $h_\infty(t)$.

The specific enthalpy $h_\mathrm{L}$ incorporates the kinematic boundary conditions at the gas-liquid interface. 
Considering a Newtonian liquid and a clean gas-liquid interface without surface-active substances, the liquid pressure $p_\mathrm{L}$ at the gas-liquid interface is defined as \citep{Ilinskii2012}
\begin{equation}
    p_{\mathrm{L}} = p_\text{G} - \frac{\alpha \sigma}{R} - 2  \alpha \,  \frac{\mu \dot{R}}{R}, \label{eq:pL}
\end{equation}
where $\sigma$ is the surface tension coefficient and $\mu$ is the dynamic viscosity of the liquid. The Gilmore model has also been used in conjunction with viscoelastic materials surrounding the gas bubble \citep{Zilonova2018,Denner2023a,Qin2023a,Li2022a} and considering a coated gas-liquid interface \citep{Guemmer2021,Denner2023,Li2022a,Qin2023a}. 

The gas pressure inside the bubble, $p_\text{G}$, is defined by an appropriate EoS as a function of the bubble size. Assuming the gas to be ideal, the pressure in the bubble to be spatially uniform and the compression to be adiabatic, the gas pressure of a cavitation bubble under consideration of the dimensionality parameter $\alpha$ is given as
\begin{equation}
    p_\text{G} = p_{\text{G},0} \left(\frac{R_0^{\alpha+1}}{R^{\alpha+1}} \right)^\gamma,
\end{equation}
where $p_{\text{G},0}$ is the gas pressure at initial radius $R_0$ and $\gamma$ is the ratio of specific heats (or, more generally, the polytropic exponent) of the gas. 
The finite volume occupied by the gas molecules, which becomes important during a strong bubble collapse, can be taken into account by the van-der-Waals hardcore radius $r_\text{hc}$,\citep{Brenner2002} such that
\begin{equation}
    p_\text{G} = p_{\text{G},0} \left(\frac{R_0^{\alpha+1}-r_\text{hc}^{\alpha+1}}{R^{\alpha+1} - r_\text{hc}^{\alpha+1}} \right)^{\gamma}.
\end{equation}
Alternatively, the same can be achieved with the Noble-Abel {(NA)} EoS \citep{Johnston2005,Toro2009}, a simplification of the NASG EoS discussed in Section \ref{sec:eos}, with which the gas pressure and density as a result of adiabatic compression are defined as \citep{Denner2021}
\begin{eqnarray}
    p_\text{G} &=& p_{\text{G},0} \left[\frac{\rho_\text{G} (1-b \, \rho_\text{G,0})}{\rho_{\text{G},0} (1-b \, \rho_\text{G})} \right]^\gamma \label{eq:pG_NA} \\
    \rho_\text{G} &=& \rho_\text{G,0} \left(\frac{R_0}{R}\right)^{\alpha+1}, \label{eq:rhoG_NA}
\end{eqnarray}
respectively, where {$\rho_\text{G,0}$ is the gas density at pressure $p_{\text{G},0}$ and} $b$ is the co-volume of the gas. More elaborate gas models that dispense with the polytropic relation or use more complex gas equations of state \citep{Prosperetti1988,Qin2023,Nazari-Mahroo2020}, account for heat transfer \citep{Prosperetti1991,Stricker2011,Zhou2020,Zhou2021a,Samiei2010}, mass transfer \citep{deGraaf2014,Byun2004,Samiei2010,Qin2023,Nazari-Mahroo2020}, and external damping effects \citep{deGraaf2014} can also be readily applied in conjunction with the Gilmore model. For underwater explosions, the Jones-Wilkins-Lee (JWL) EoS \citep{Lee1968,Baudin2010} is a popular choice in conjunction with the Chapman-Jouguet isentrope \citep{Chapman1899,Jouguet1905} to describe the properties of the gaseous detonation products constituting the bubble and to provide initial conditions for the velocity of the gas-liquid interface \citep{Riley2010,Kwak2012,Zhang2021c,Zhang2021f,Wang2021c,Gao2022b,Jia2022a,Sheng2023}.

The Gilmore equation for spherical bubbles, Eq.~\eqref{eq:bubbleODE_gilmore}, has been shown in many studies to yield remarkably accurate results, which Prosperetti and Lezzi \citep{Prosperetti1986,Lezzi1987} attributed to the fact that the specific enthalpy of the liquid is retained in the equation of motion. Even though the Gilmore equation contains some high-order terms, it is not consistently second-order accurate with respect to the Mach number of the gas-liquid interface. \citet{Benjamin1958} pointed to missing second-order terms in the Kirkwood-Bethe hypothesis, which was met with some skepticism by Plesset (see the discussion on p.~232-233 in the book edited by \citet{Cooper1958}), although similar second-order errors in the Gilmore model \citep{Gilmore1952} were later identified by \citet{Tilmann1980} and \citet{Prosperetti1986}. \citet{Lezzi1987} proposed a more elaborate equation of motion for pressure-driven dynamics of spherical bubbles that is second-order accurate with respect to the Mach number, which was, however, derived under the explicit assumption that the velocity of the induced flow is small. Recent studies \citep{Zhang2021c,Wang2021c,Sheng2023} have used the model of \citet{Lezzi1987} in conjunction with the Kirkwood-Bethe hypothesis successfully to predict the pressure transients and shock waves generated by underwater explosions.

\section{Flow field around a bubble}
\label{sec:liquid}

The flow field generated by an oscillating or collapsing bubble, or an underwater explosion, can be predicted with expressions derived from the Kirkwood-Bethe hypothesis in conjunction with the appropriate conservation laws. To this end, the original work of \citet{Kirkwood1942}, as well as subsequent studies by \citet{Rice1943,Rice1944}, \citet{Akulichev1968}, and \citet{Kedrinskii1972} focused on deriving models for the evolution of the peak pressure of a shock wave emitted by an underwater explosion or a pressure-driven bubble, leveraging the Kirkwood-Bethe hypothesis together with the Rankine-Hugoniot conditions. Since the Rankine-Hugoniot conditions follow from the governing conservation laws at shock discontinuities, these peak-pressure models can be seen a special case of the more general descriptions of the flow field devised by \citet{Gilmore1952}, \citet{Hickling1963}, and \citet{Ivany1965}. 

The enthalpy of the fluid is readily obtained by rearranging Eq.~\eqref{eq:kb_general} as
\begin{equation}
    h = h_\infty + \frac{g}{r^{\alpha/2}} - \frac{u^2}{2} , \label{eq:h_KB}
\end{equation}
where the invariant $g$ is defined by Eq.~\eqref{eq:kb_general}, using the conditions under which it originates at the gas-liquid interface ($R=r$), as
\begin{equation}
    g = R^{\alpha/2} \, \Omega_\text{L} = R^{\alpha/2} \left(h_\text{L}- h_\infty + \frac{\dot{R}^2}{2}\right).
    \label{eq:kb_general_wall}
\end{equation}
Inserting Eq.~\eqref{eq:kb_general_wall} into Eq.~\eqref{eq:h_KB}, the specific enthalpy
at radial location $r$ and time $t$ is
\begin{equation}
    h(r,t) = h_\infty(t) + \left[\frac{R(\tau)}{r(t)} \right]^{\alpha/2} \Omega_\text{L}(\tau) - \frac{u(r,t)^2}{2} , \label{eq:h_KB_full}
\end{equation}
where $\tau = t-r/(c+u)$ is the retarded time that corresponds to the time at which the  invariant $g$ or, likewise, the specific kinetic enthalpy $\Omega$ are emitted at the gas-liquid interface. This definition of the specific enthalpy satisfies the boundary condition $h(R,\tau)=h_\text{L}(\tau)$ at the gas-liquid interface ($r=R$). Considering, in addition, that $h_\infty$ is an explicitly imposed ambient enthalpy and the radial position $r$ can be obtained by integrating Eq.~\eqref{eq:drdt_char}, the flow velocity $u$ is the only remaining unknown. 

Eqs.~\eqref{eq:drdt_char} and \eqref{eq:h_KB_full}, in conjunction with either the explicit expression, Eq.~\eqref{eq:u_expl_expanded}, or the differential equation, Eq.~\eqref{eq:dudt_char_with_drdt}, for the flow velocity derived below, describe the flow field induced by a bubble or cavity with varying size, assuming the induced flow is isentropic.

\subsection{Explicit expression for the flow velocity}
\label{sec:liquid_expl}

By directly inserting the velocity potential $\psi = \phi/r^{\alpha/2}$ into the definition of the velocity given by Eq.~\eqref{eq:u_potential}, similar to the derivation of the quasi-acoustic description of the flow field presented by \citet{Gilmore1952} and \citet{Trilling1952}, we obtain 
\begin{equation}
    u = \frac{\alpha \, \phi}{2 \, r^{1+\alpha/2}} - \frac{1}{r^{\alpha/2}} \frac{\partial \phi}{\partial r}. \label{eq:u_expl_dphidr}
\end{equation}
Making the assumption that the potential $\phi$ propagates with constant value along outgoing characteristics, 
\begin{equation}
    \left. \frac{\text{d}\phi}{\text{d}t}\right|_\text{c} = \frac{\partial \phi}{\partial t} + \left. \frac{\text{d}r}{\text{d}t}\right|_\text{c} \frac{\partial \phi}{\partial r} =0,
\end{equation}
allows us to write
\begin{equation}
    \frac{\partial \phi}{\partial r} = - \left. \frac{\text{d}r}{\text{d}t}\right|_\text{c}^{-1} \frac{\partial \phi}{\partial t}.
\end{equation}
Inserting this expression into Eq.~\eqref{eq:u_expl_dphidr} with the propagation velocity given by Eq.~\eqref{eq:drdt_char} and  $g = \partial \phi/\partial t$, the flow velocity is defined as
\begin{equation}
    u = \frac{\alpha \, \phi}{2 \, r^{1+\alpha/2}} + \frac{g}{r^{\alpha/2} (c+u)}. \label{eq:u_expl}
\end{equation}
This explicit expression of the flow velocity can already be found in the original report of \citet{Kirkwood1942} for spherical symmetry ($\alpha=2$) as well as in the report of \citet{Rice1944} for cylindrical symmetry ($\alpha=1$). Evaluating Eq.~\eqref{eq:u_expl} at the gas-liquid interface ($R=r$), we find the invariant potential $\phi$ to be defined as
\begin{eqnarray}
    \phi = \frac{2}{\alpha} \left( R^{1+\alpha/2} \dot{R} - R \, \frac{g}{c_\text{L}+\dot{R}} \right). \label{eq:f_general_wall}
\end{eqnarray}
Inserting Eqs.~\eqref{eq:kb_general_wall} and \eqref{eq:f_general_wall} into Eq.~\eqref{eq:u_expl}, the velocity at radial location $r$ and time $t$ is
\begin{multline}
    u(r,t) = \left[\frac{R(\tau)}{r(t)}\right]^{1+\alpha/2} \left[\dot{R}(\tau) - \frac{\Omega_\text{L} (\tau)}{c_\text{L}(\tau) + \dot{R}(\tau)} \right]\\ + \left[\frac{R(\tau)}{r(t)}\right]^{\alpha/2} \frac{\Omega_\text{L}(\tau)}{c(r,t)+u(r,t)}, \label{eq:u_expl_expanded}
\end{multline}
which satisfies the boundary condition $u(R,\tau) = \dot{R}(\tau)$ at the gas-liquid interface ($r=R$). 

The explicit expression of the velocity given by Eq.~\eqref{eq:u_expl_expanded} separates the flow velocity into three distinct contributions \citep{Denner2023}: the first term is the incompressible displacement of the fluid by the moving gas-liquid interface, the second term represents the compression/expansion of the fluid resulting from its displacement, and the third term is the particle velocity associated with the emitted acoustic wave. Hence, the first two terms represent hydrodynamic contributions and the third term represents the acoustic contribution to the flow field. However, the assumption that the potential $\phi$ propagates as an outgoing wave with constant amplitude is valid only if the nonlinear terms on the right-hand side of Eq.~\eqref{eq:kb_wave_general} are negligible. Hence, Eq.~\eqref{eq:u_expl} is accurate only if the flow velocity is sufficiently small, $u^2 \ll c^2 $.

\subsection{Differential equation for the flow velocity}

Alternative to the explicit expression for the flow velocity presented in the previous section, it is also possible to derive an ODE describing the flow velocity, as proposed by \citet{Gilmore1952} and \citet{Hickling1963}, without the assumption that $\phi$ retains a constant value as it propagates outward. To this end, the continuity equation, Eq.~\eqref{eq:continuity_h}, is added to the momentum equation, Eq.~\eqref{eq:momentum_h}, to obtain
\begin{multline}
    \frac{1}{c}\left( \frac{\partial h}{\partial t} + u \frac{\partial h}{\partial r}\right) + c \frac{\partial u}{\partial r} + \frac{\alpha u c}{r}  +
    \frac{\partial u}{\partial t} \\ + u \frac{\partial u}{\partial r} + \frac{\partial h}{\partial r}  = \frac{1}{c} \, \frac{\partial h_\infty}{\partial t}. \label{eq:continuity-momentum}
\end{multline}
This equation may be rearranged as
\begin{multline}
    \frac{1}{c}\left[ \frac{\partial h}{\partial t} + (c+u) \frac{\partial h}{\partial r}\right]  +
    \frac{\partial u}{\partial t} + (c+u) \frac{\partial u}{\partial r} \\ = - \frac{\alpha u c}{r} + \frac{1}{c} \, \frac{\partial h_\infty}{\partial t}, \label{eq:continuity-momentum2}
\end{multline}
which is, considering that the ambient specific enthalpy $h_\infty$ is assumed to be spatially invariant, equivalent to the equation for the outgoing Riemann invariant, see Eq.~\eqref{eq:riemann_outgoing}. Thus, the derivation of a differential equation for the flow velocity starts from the equation governing a curved Riemann wave in an isentropic flow. This further shows that the propagation speed $c+u$, albeit only approximately accurate, is inferred by the governing conservation laws. 

The enthalpy derivative along outgoing characteristics follows from Eq.~\eqref{eq:continuity-momentum2} as 
\begin{eqnarray}
     \left. \frac{\mathrm{d}h}{\mathrm{d}t}\right|_\text{c} =- c  \left. \frac{\mathrm{d}u}{\mathrm{d}t}\right|_\text{c} - \frac{\alpha u c^2}{r} + \frac{\partial h_\infty}{\partial t}.
     \label{eq:dhdt_char}
\end{eqnarray}
Taking the derivative of Eq.~\eqref{eq:kb_general} along an outgoing characteristic reads as
\begin{multline}
      \left. \frac{\mathrm{d}g}{\mathrm{d}t}\right|_\mathrm{c} = \frac{\alpha g}{2 r}  \left. \frac{\mathrm{d}r}{\mathrm{d}t}\right|_\mathrm{c} + r^{\alpha/2} \left. \frac{\mathrm{d}h}{\mathrm{d}t}\right|_\mathrm{c} \\ - r^{\alpha/2} \left. \frac{\mathrm{d}h_\infty}{\mathrm{d}t}\right|_\mathrm{c}  +  r^{\alpha/2} u  \left. \frac{\mathrm{d}u}{\mathrm{d}t}\right|_\mathrm{c} =0,  \label{eq:dhdt_char_gbased}
\end{multline}
where $(\mathrm{d}g/\mathrm{d}t)_\mathrm{c} = 0$ by virtue of the Kirkwood-Bethe hypothesis, and $(\mathrm{d}h_\infty/\mathrm{d}t)_\mathrm{c} = \partial h_\infty / \partial t $, since $h_\infty$ is spatially invariant by definition. Inserting Eq.~\eqref{eq:dhdt_char_gbased} into Eq.~\eqref{eq:dhdt_char},
\begin{multline}
     \frac{\partial h_\infty}{\partial t} - \frac{\alpha g}{2 r^{(1+\alpha/2)}} \left. \frac{\mathrm{d}r}{\mathrm{d}t}\right|_\mathrm{c}  - u  \left. \frac{\mathrm{d}u}{\mathrm{d}t}\right|_\mathrm{c}  \\ =- c \left. \frac{\mathrm{d}u}{\mathrm{d}t}\right|_c - \frac{\alpha u c^2}{r} +\frac{\partial h_\infty}{\partial t} , 
\end{multline}
the derivative of the flow velocity $u$ along an outgoing characteristic is given as
\begin{equation}
    \left. \frac{\mathrm{d}u}{\mathrm{d}t}\right|_c = \frac{\alpha}{r(c-u)} \bigg[ \frac{g}{2 \, r^{\alpha/2}} \left. \frac{\mathrm{d}r}{\mathrm{d}t}\right|_\mathrm{c} -  u c^2 \bigg],
    \label{eq:dudt_char_with_drdt}
\end{equation}
or, by inserting Eq.~\eqref{eq:drdt_char}, 
\begin{equation}
    \left. \frac{\mathrm{d}u}{\mathrm{d}t}\right|_c = \frac{\alpha}{r(c-u)} \bigg[ \frac{g}{2 \, r^{\alpha/2}} \, (c+u) -  u c^2 \bigg].
    \label{eq:dudt_char}
\end{equation}
In spherical symmetry ($\alpha=2$), Eq.~\eqref{eq:dudt_char} corresponds to the ODE first presented by \citet{Hickling1963},
\begin{equation}
    \left. \frac{\mathrm{d}u}{\mathrm{d}t}\right|_c = \frac{1}{r(c-u)} \left[\frac{g}{r}\, (c+u) - 2 u c^2 \right].
    \label{eq:dudt_char_sphere}
\end{equation}
The derivation of this ODE for the flow velocity in the way presented here follows the derivation presented for the special case of spherical symmetry ($\alpha=2$) by \citet{Ivany1965}.

An ODE for the flow velocity with respect to the radial coordinate may be obtained by dividing Eq.~\eqref{eq:dudt_char} by Eq.~\eqref{eq:drdt_char},
\begin{equation}
    \left. \frac{\mathrm{d}u}{\mathrm{d}r}\right|_c =  \left. \frac{\mathrm{d}u}{\mathrm{d}t}\right|_c  \left. \frac{\mathrm{d}r}{\mathrm{d}t}\right|_c^{-1} = \frac{\alpha}{r(c-u)} \bigg[ \frac{g}{2 \, r^{\alpha/2}} -  \frac{u c^2}{c+u} \bigg],
    \label{eq:dudt_char_Gilmore}
\end{equation}
an ODE first proposed by \citet{Gilmore1952} for spherical symmetry ($\alpha=2$) using the modified Tait EoS.

\subsection{Liquid pressure}

With an EoS for the liquid at hand, the pressure of the liquid can be computed based on the specific enthalpy $h$. Using the modified Tait EoS, the pressure of the liquid follows by inserting Eq.~\eqref{eq:rho_Tait} into Eq.~\eqref{eq:h_Tait} to obtain
\begin{equation}
    p = \left[ \frac{(n-1) \rho_0}{n \left( p_0 + B \right)^{1/n}} \, h \right]^{\frac{1}{1-1/n}} - B \label{eq:p_Tait}
\end{equation}
or, using Eqs.~\eqref{eq:h_Tait} and \eqref{eq:h_KB}, the pressure can be computed based on the invariant $g$,
\begin{equation}
    p = \rho \left[ \frac{p_\infty + B}{\rho_\infty}  + \frac{n-1}{n}  \left(\frac{g}{r^{\alpha/2}} - \frac{u^2}{2}\right)\right] - B, \label{eq:p_Tait_Ebeling}
\end{equation}
as introduced for spherical symmetry ($\alpha=2$) by \citet{Ivany1965} and, later, \citet{Ebeling1978}. The pressure amplitude $\Delta p = p - p_\infty$ is dominated by the term $g/r^{\alpha/2}$, especially in the far field where $u\rightarrow 0$, thus recovering the well-known amplitude decay with $1/r$ in spherical symmetry ($\alpha=2$) and $1/\sqrt{r}$ in cylindrical symmetry ($\alpha=1$).\citep{Landau1959} 
With the NASG EoS, the liquid pressure follows from Eqs.~\eqref{eq:h_NASG} and \eqref{eq:h_KB} in a similar manner as
\begin{equation}
   p = \frac{(n-1) \rho h - (1-b \rho) n B}{n - b \rho}.
   \label{eq:p_NASG_Denner}
\end{equation}
Since $p$ and $\rho$ depend explicitly on each other, Eqs.~\eqref{eq:p_Tait_Ebeling} and \eqref{eq:p_NASG_Denner} may require an iterative evaluation \citep{Denner2023}, in order to reach a desired accuracy.

Alternatively, inserting Eq.~\eqref{eq:dudt_char} into Eq.~\eqref{eq:dhdt_char_gbased}, and rearranging for the enthalpy derivative yields
\begin{equation}
    \left. \frac{\text{d}h}{\text{d}t} \right|_\text{c} = \frac{\alpha}{r(c-u)} \left[u^2 c^2 - \frac{g}{2 r^{\alpha/2}} (c+u) c \right] + \frac{\partial h_\infty}{\partial t}.
\end{equation}
By applying Eq.~\eqref{eq:dh_dp_drho}, the ODE for the pressure along outgoing characteristics follows as
\begin{equation}
    \left. \frac{\text{d}p}{\text{d}t} \right|_\text{c} = \frac{ \rho \alpha}{r(c-u)} \left[u^2 c^2 - \frac{g}{2 r^{\alpha/2}} (c+u) c \right] +  \frac{\partial p_\infty}{\partial t},
    \label{eq:dpdt_char}
\end{equation}
which may be solved together with the ODEs for the radial position, Eq.~\eqref{eq:drdt_char}, and the velocity, Eq.~\eqref{eq:dudt_char}. Assuming spherical symmetry in conjunction with the modified Tait EoS, and neglecting ambient pressure changes, Eq.~\eqref{eq:dpdt_char} simplifies to
\begin{equation}
    \left. \frac{\text{d}p}{\text{d}t} \right|_\text{c} = \frac{\rho_0}{r(c-u)} \left(\frac{p+B}{p_0+B}\right)^{1/n} \left[2 u^2 c^2 - \frac{g}{r} (c+u) c \right],
    \label{eq:dpdt_char_Tait}
\end{equation}
which has been used frequently instead of Eq.~\eqref{eq:p_Tait_Ebeling}.

\section{Model hierarchy}
\label{sec:hierarchy}

KBH models for pressure-driven bubble dynamics and acoustics are part of a consistent hierarchy of models \citep{Naugolnykh1971,Prosperetti1984a,Denner2023,Zhang2023b} that can be established by successive simplifications of the Kirkwood-Bethe hypothesis.

Assuming $u^2\ll c_0^2$ and constant fluid properties ($\rho=\rho_0$, $c=c_0$), such that the EoS of the liquid becomes obsolete, the potential $\psi = \phi/ r^{\alpha/2}$ is governed by the wave equation 
\begin{equation}
    \frac{1}{c_0^2} \frac{\partial^2 \psi}{\partial t^2} - \nabla^2 \psi = 0,
\end{equation}
the propagation speed along outgoing characteristics is constant,
\begin{equation}
    \left. \frac{\text{d}r}{\text{d}t} \right|_\text{c} = c_0, \label{eq:drdt_char_qa}
\end{equation}
and the specific enthalpy is defined by the first-order approximation $h = p/\rho_0$. The pressure, consequently, follows from Eq.~\eqref{eq:h_KB} as
\begin{equation}
    p = p_\infty + \rho_0 \left(\frac{g}{r^{\alpha/2}} - \frac{u^2}{2}\right), \label{eq:p_qa}
\end{equation}
which neglects terms proportional to $c_0^{-2}$ and higher. Considering Eq.~\eqref{eq:drdt_char_qa}, the explicit expression for the flow velocity, Eq.~\eqref{eq:u_expl}, simplifies to
\begin{equation}
    u = \frac{\alpha \, \phi}{2 \, r^{1+\alpha/2}} + \frac{g}{r^{\alpha/2} c_0}. \label{eq:u_expl_qa}
\end{equation}
Based on Eqs.~\eqref{eq:p_qa} and \eqref{eq:u_expl_qa}, the invariants $g$ and $\phi$ are defined at the gas-liquid interface ($r=R$), such that the boundary conditions are satisfied, as
\begin{eqnarray}
    g &=& R^{\alpha/2} \left(\frac{p_\text{L}-p_\infty}{\rho_0} + \frac{\dot{R}^2}{2}\right)\\
    \phi &=& \frac{2}{\alpha} \left( R^{1+\alpha/2} \dot{R} - R \, \frac{g}{c_0} \right).
\end{eqnarray}
This presents, in fact, the quasi-acoustic model of \citet{Gilmore1952} and \citet{Trilling1952}, generalized with respect to the dimensionality coefficient $\alpha$. Although this quasi-acoustic model incorporates a finite wave speed, it cannot describe the nonlinear distortion of acoustic waves and the formation of shock fronts, since all parts of a wave propagate with the same constant speed $c_0$. Applying the same simplifications to the ODE for the radial bubble dynamics, Eq.~\eqref{eq:bubbleODE}, and neglecting the term proportional to $c_0^{-2}$, yields a generalized form of the Keller-Miksis equation,\citep{Keller1980}
\begin{multline}
    \left(1 - \frac{\dot{R}}{c_0}\right) R \ddot{R} + \frac{3}{4} \alpha \left(1 - \frac{\dot{R}}{3\, c_0}\right) \dot{R}^2  \\= \frac{\alpha}{2} \left(1 + \frac{\dot{R}}{c_0}\right) \frac{p_\mathrm{L} - p_\infty}{\rho_0} + R \, \frac{\dot{p}_\mathrm{L}-\dot{p}_\infty}{\rho_0 \, c_0} .
    \label{eq:km}
\end{multline}
The quasi-acoustic model and the Keller-Miksis equation are first-order accurate with respect to the Mach number and, thus, limited to small Mach numbers, $(\dot{R}/c_0)^2 \ll 1$.

Going one step further by assuming the fluid is incompressible, with $c_0 \rightarrow \infty$, the velocity potential is governed by $\nabla^2 \psi = 0$ and the propagation speed along outgoing characteristics is infinite.
The pressure in the liquid is given by Eq.~\eqref{eq:p_qa} and the flow velocity follows from Eq.~\eqref{eq:u_expl_qa} as
\begin{equation}
    u = \left( \frac{R}{r} \right)^{1+\alpha/2} \dot{R}.
\end{equation}
With $c_0\rightarrow \infty$, the ODE for the radial bubble dynamics, Eq.~\eqref{eq:km}, further reduces to a generalized form of the Rayleigh-Plesset equation \citep{Plesset1977,Ilinskii2012,Wang2024c},
\begin{equation}
    R \ddot{R} + \frac{3}{4} \alpha \dot{R}^2 =\frac{\alpha (p_\mathrm{L} - p_\infty)}{2 \rho_0}.
    \label{eq:standardRP}
\end{equation}

\section{Solution methods}
\label{sec:numerics}

The equations governing the bubble dynamics and the equations describing the flow field can be solved independent of each other. From the viewpoint of the bubble dynamics, this is possible because the Kirkwood-Bethe hypothesis connects the kinematic conditions at the gas-liquid interface to the spatially invariant ambient condition at an assumed infinite distance. The flow field, in turn, is defined by discrete pieces of information that evolve along a single spatial dimension with speed $c+u$ and are connected to their initial conditions originating at the gas-liquid interface by the invariant $g$. 
{The solution methods presented in this section are implemented in the open-source software library {\tt APECSS} \citep{Denner2023a}.}

\subsection{Bubble dynamics}

In order to solve the equation of motion of the gas-liquid interface, Eq.~\eqref{eq:bubbleODE}, we require expressions for the specific enthalpy difference $H$ and its derivative $\dot{H}$. The specific enthalpy difference can be computed using a suitable EoS in conjunction with the liquid pressure at the gas-liquid interface $p_\text{L}$ and the ambient pressure $p_\infty$. 

Considering a Newtonian liquid, $p_\text{L}$ is given by Eq.~\eqref{eq:pL}. Since $\text{d}h = \text{d}p/\rho$ for the considered isentropic flow, the derivative of the specific enthalpy difference is defined as
\begin{multline}
    \dot{H} = \frac{\dot{p}_\text{L}}{\rho_\text{L}} - \frac{\dot{p}_\infty}{\rho_\infty} = \frac{1}{\rho_\text{L}} \left(\dot{p}_\text{G} + \frac{\alpha \sigma}{R^2} \dot{R} \right. \\ \left. + 2 \alpha \mu \frac{\dot{R}^2}{R^2} - 2 \alpha \mu \frac{\ddot{R}}{R} \right) - \frac{\dot{p}_\infty}{\rho_\infty}. \label{eq:Hdot}
\end{multline}
Inserting Eq.~\eqref{eq:Hdot} into Eq.~\eqref{eq:bubbleODE} and gathering the $\ddot{R}$-terms on the left-hand side yields 
\begin{multline}
    \left(c_\text{L} R + \frac{2 \alpha \mu}{\rho_\text{L}} \right) \ddot{R} = \frac{\alpha}{2} \left(\frac{ c_\text{L} + {\dot{R}}}{c_\text{L} - {\dot{R}}} \, H -  \frac{3 c_\text{L} - {\dot{R}}}{c_\text{L} - {\dot{R}}} \, \frac{\dot{R}^2}{2}\right) + \\
    \frac{R}{\rho_\text{L}} \left(\dot{p}_\text{G} + \frac{\alpha \sigma}{R^2} \dot{R} + 2 \alpha \mu \frac{\dot{R}^2}{R^2}  \right) - 
    \frac{R \dot{p}_\infty}{\rho_\infty }. \label{eq:bubbleODE_lhs}
\end{multline} 
We can solve this second-order ODE by separating it into two first-order ODEs for $\dot{R}$ and $\dot{U}=\ddot{R}$, such that at each time instance $t_j$ we integrate the equations
\begin{eqnarray}
  \dot{R}_{j+1} = U_j \label{eq:bubbleODE_R}
\end{eqnarray}
and
\begin{multline}
   \dot{U}_{j+1} = \frac{\alpha}{2 \mathcal{A}_j} \left(\frac{ c_{\text{L},j} + {\dot{R}_j}}{c_{\text{L},j} - {\dot{R}_j}}  H_j -  \frac{3 c_{\text{L},j} - {\dot{R}_j}}{c_{\text{L},j} - {\dot{R}_j}} \, \frac{\dot{R}_j^2}{2}\right) \\ + 
    \frac{R_j}{\mathcal{A}_j} \left(\frac{\dot{p}_{\text{G},j}}{\rho_{\text{L},j}} + \frac{\alpha \sigma \dot{R}_j}{R_j^2 \rho_{\text{L},j}}  + \frac{2 \alpha \mu \dot{R}_j^2}{R^2_j \rho_{\text{L},j}}   - 
    \frac{\dot{p}_{\infty,j}}{\rho_{\infty,j} }\right) , \label{eq:bubbleODE_U}
\end{multline}
where
\begin{equation}
    \mathcal{A}_j = c_{\text{L},j} R_j + \frac{2 \alpha \mu}{\rho_{\text{L},j}} .
\end{equation}

The embedded fifth-order Runge-Kutta scheme of \citet{Dormand1980}, which adapts the numerical time step $\Delta t_j = t_j - t_{j-1}$ based on the estimated solution error, is a common choice to integrate this system of ODEs, as it is, for instance, the basis of the {\tt ode45} function in Matlab, the default solver of the {\tt solve\_ivp} function of the SciPy package, and used in the solver of the open-source cavitation software library {\tt APECSS} \citep{Denner2023a}.

\subsection{Flow field}
\label{sec:numerics_flow}

To solve the flow velocity and specific enthalpy (or pressure), the information originating at the gas-liquid interface is tracked along the outgoing characteristics, whereby each parcel of information propagates on its own characteristic with velocity $c+u$. To this end, the bubble radius $R_j$, the speed of the gas-liquid interface $\dot{R}_j$, and the specific enthalpy of the liquid at the gas-liquid interface $h_{\text{L},j}$ obtained by solving Eqs.~\eqref{eq:bubbleODE_R} and \eqref{eq:bubbleODE_U} serve as the initial conditions.

The invariant $g$ of information parcel $i$ is defined upon emission at the gas-liquid interface, at emission time $\tau_i$, as
\begin{equation}
    g_{i}  = R(\tau_i)^{\alpha/2}  \left[ h_{\text{L}}(\tau_i) - h_{\infty}(\tau_i) + \frac{\dot{R}(\tau_i)^2}{2} \right]. \label{eq:g_disc}
\end{equation}
The radial position $r_{i,j}$ and flow velocity $u_{i,j}$ of each information parcel $i$ at time $t_j$ result from Eqs.~\eqref{eq:drdt_char} and \eqref{eq:dudt_char_with_drdt}, by integrating
\begin{eqnarray}
  \left. \frac{\mathrm{d}r}{\mathrm{d}t}\right|_{\text{c}(i,j)} = c_{i,j} + u_{i,j} \label{eq:drdt_char_disc}
\end{eqnarray}
\begin{multline}
    \left. \frac{\mathrm{d}u}{\mathrm{d}t}\right|_{\text{c}(i,j)} =  \frac{\alpha}{r_{i,j} (c_{i,j}-u_{i,j})} \Bigg[ \frac{g_{i}}{2 \, r_{i,j}^{\alpha/2}} \left. \frac{\mathrm{d}r}{\mathrm{d}t}\right|_{\mathrm{c}(i,j)} -  u_{i,j} c_{i,j}^2 \Bigg]  \label{eq:dudt_char_disc}
\end{multline}
using a suitable integration scheme for ODEs together with the initial conditions $R(\tau_i)$ and $\dot{R}(\tau_i)$. Once the new radial position $r_{i,j}$ and flow velocity $u_{i,j}$ of information parcel $i$ are determined, its specific enthalpy is readily given by Eq.~\eqref{eq:h_KB} as
\begin{equation}
    h_{i,j} = h_{\infty,j} + \frac{g_{i}}{r_{i,j}^{\alpha/2}} - \frac{u_{i,j}^2}{2}, \label{eq:h_KB_disc}
\end{equation}
from which the local pressure and speed of sound can be computed using the employed EoS. 

As an alternative to integrating the velocity using Eq.~\eqref{eq:dudt_char_disc}, although not used for the results presented in Section \ref{sec:results}, the flow velocity may also be determined explicitly using Eq.~\eqref{eq:u_expl_expanded}, such that 
\begin{equation}
    u_{i,j} = \frac{\alpha \, \phi_i}{2 \, r_{i,j}^{1+\alpha/2}} + \frac{g_i}{r_{i,j}^{\alpha/2} (c_{i,j}+u_{i,j})}, \label{eq:u_expl_disc}
\end{equation}
with 
\begin{eqnarray}
    \phi_i = \frac{2}{\alpha} \left[ R(\tau_i)^{1+\alpha/2} \dot{R}(\tau_i) - R(\tau_i) \, \frac{g_i}{c_\text{L}(\tau_i)+\dot{R}(\tau_i)} \right]  \label{eq:f_disc}
\end{eqnarray}
if sufficiently small flow velocities are expected.

\citet{Ivany1965} proposed to solve the flow velocity and pressure along outgoing characteristics originating at the gas-liquid interface at selected time instances, after the solution of the bubble dynamics had been fully obtained. \citet{Denner2023} presented a Lagrangian wave tracking algorithm, whereby the flow field is solved together with the bubble dynamics and the information originating from the gas-liquid interface is tracked along outgoing characteristics. To this end, Eqs.~\eqref{eq:drdt_char_disc} and \eqref{eq:dudt_char_disc} are integrated together as a coupled system of ODEs by a conventional fourth-order Runge-Kutta scheme, using the same time step $\Delta t_j$ as for the integration of Eqs.~\eqref{eq:bubbleODE_R} and \eqref{eq:bubbleODE_U} \citep{Denner2023a}. 

\subsection{Shock fronts}
\label{sec:numerics_shocks}

Because both the particle velocity and the speed of sound are monotone-increasing functions of the pressure \citep{Fay1931,Hamilton1988}, the parts of an acoustic wave with a higher pressure propagate faster than the parts of the same wave with a lower pressure. This pressure dependency of the propagation speed drives a progressive steepening of the emitted waves and, eventually, may lead to the formation of a shock front for waves with sufficiently large initial amplitude. These cumulative nonlinear effects may become dominant for $Mkr \gtrsim 1$, {where $k$ is the wavenumber of the emitted wave,} irrespective of the Mach number $M$; \citep{Naugolnykh1971a} in the absence of dissipation, any plane wave forms a shock wave in finite time. At the shock front, a frequency-dependent attenuation prevents fluid particles from overtaking the shock front \citep{Fay1931} and, as postulated by \citet{Rudnick1952}, the stabilization of the shock front is independent of the origin of the acting dissipation process, such as viscous stresses or thermal conduction. This can be observed in numerical simulations, where a stable shock front can form even in inviscid and non-conducting flows, if sufficient numerical dissipation is present \citep{Laney1998,LeVeque2008,Schenke2022}.

\begin{figure*}
    \includegraphics[width=0.85\linewidth]{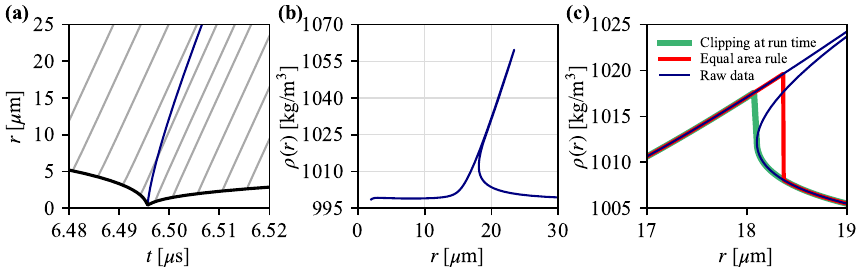}
        \caption{Example of the different treatments of a shock front forming after the collapse of a cavitation bubble produced by a femtosecond laser, as studied by \citet{Liang2022}. (a) Selected outgoing wave characteristics are shown in a time-space plot, where the characteristic associated with the minimum radius is seen to intersect previously emitted characteristics. The thick black line shows the bubble radius. (b) The resulting multivalued density profile $50 \, \text{ns}$ after the bubble attained its minimum radius. (c) Close up of the density profile, to which two different shock treatments are applied. The clipping of the profile as proposed by \citet{Denner2023} is applied at run time, whereas the rule of equal areas \citep{Landau1959} is applied as a post-processing step.}
        \label{fig:shock_sketch}
\end{figure*}
    
When solving the flow field by propagating the relevant information along outgoing characteristics using a KBH model, the formation of a shock front is signified by the intersection of characteristics, as seen in Figure \ref{fig:shock_sketch}(a). This intersection of characteristics yields a multivalued solution, shown in Figure \ref{fig:shock_sketch}(b), that is not physical. \citet{Akulichev1968} applied the \textit{rule of equal areas} \citep{Landau1959} to post process the multivalued solutions resulting from intersecting characteristics. Taking a multivalued density profile as shown in Figure \ref{fig:shock_sketch}(b), the corresponding weak solution is obtained graphically by eliminating the multivalued solution in such a way, that the area under the curve remains unchanged, as illustrated in Figure \ref{fig:shock_sketch}(c). Thus, the integral of the multivalued solution and the integral of the discontinuous weak solution of the density profile are the same, such that mass is conserved \citep{Whitham1999}. This implies that the wave on either side of the shock front remains a simple wave traveling in one direction, without reflections introduced at the discontinuity \citep{Naugolnykh1971a}. This method has been used, for instance, in studies on laser-induced cavitation \citep{Vogel1996,Lai2022,Liang2022,Wen2023a} and sonoluminescence \citep{Holzfuss2010} to model the formation and attenuation of shock waves using the Kirkwood-Bethe hypothesis. 

\begin{figure}
    \centering
    \includegraphics[scale=1]{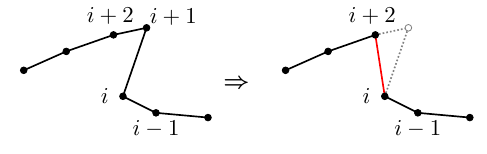}
    \caption{Sketch of a discrete wave profile (left) with a multivalued solution and (right) after treating this multivalued solution by \textit{clipping} the wave profile, as proposed by \citet{Denner2023}.}
    \label{fig:shock_clipping}
\end{figure}

Utilizing the Lagrangian wave tracking algorithm mentioned in the previous section, \citet{Denner2023} proposed to simply discard any information that overtakes the information propagating ahead of it, as illustrated in Figure \ref{fig:shock_clipping}, to account for the formation of shock fronts and the associated dissipation. \citet{Naugolnykh1971a} described a similar procedure as ``clipping of the vertices of its profile'' when referring to the decay of plane sawtooth waves. This rather simple procedure aims to exploit the argument of \citet{Rudnick1952} that the stabilization of the shock front is independent of the origin of dissipation. However, even though the results obtained with this method are in reasonably good agreement with the result obtained by applying the rule of equal areas for shock waves produced by collapsing bubbles \citep{Denner2023}, discarding information by clipping the density profile does not redistribute mass. Hence, this simple shock treatment yields an error in mass conservation and overpredicts the attenuation of the shock front, as observed in Figure \ref{fig:shock_sketch}(c).

\section{Further investigation}
\label{sec:results}

KBH models have been tested and validated extensively \citep{Oh2018,Geng2021,Vassholz2021,Lai2022,Wen2023a,Li2023b,Zhang2021f,Wang2021c,Sheng2023,Denner2023}, utilizing high-fidelity experiments and fully resolved simulations, and have been found to produce accurate and reliable results in a variety of scenarios. However, some fundamental questions regarding the validity and applicability of the Kirkwood-Bethe hypothesis remain. To further scrutinize and improve KBH models, this section focuses on the prediction of the propagation and attenuation of shock fronts at simulation run time (Section \ref{sec:results_shocks}) and the complex impedance of curved acoustic waves (Section \ref{sec:results_impedance}). 
{The presented results were produced with version v1.6 of the open-source software library {\tt APECSS} \citep{Denner2023a}, available at \url{https://doi.org/10.5281/zenodo.10981878}.}

\subsection{Predicting shock fronts at simulation run time}
\label{sec:results_shocks}

\begin{figure}
    \centering
    \includegraphics[scale=1]{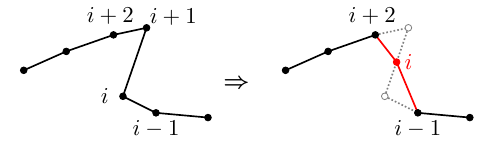}
    \caption{Sketch of a discrete wave profile (left) with a multivalued solution and (right) after treating this multivalued solution by the proposed \textit{averaging} procedure, Eqs.~\eqref{eq:r_shock_new}-\eqref{eq:u_shock_new}.}
    \label{fig:shock_averaging}
\end{figure}

\begin{figure*}
  \includegraphics[width=0.85\linewidth]{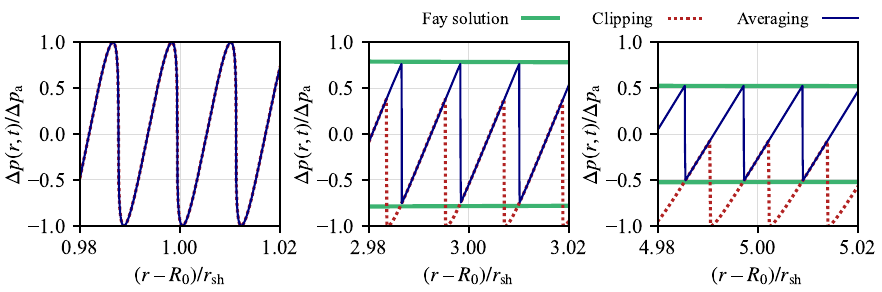}
  \caption{Pressure amplitude $\Delta p(r,t)$, normalized by the excitation pressure amplitude $\Delta p_\text{a}$,  generated by an oscillating planar emitter as a function of the dimensionless distance $(r-R_0)/r_\text{sh}$, where $R_0$ is the initial size of the emitter and $r_\text{sh}$ is the shock formation distance of a plane harmonic wave in an inviscid fluid, Eq.~\eqref{eq:shock_distance}. The analytical solution for the amplitude of the decaying shock wave of \citet{Fay1931}, Eq.~\eqref{eq:fay}, is shown for reference.}
  \label{fig:Emitter_planar}
\end{figure*}

The robust and accurate treatment of multivalued solutions associated with shock fronts at simulation run time presents challenges. 
{Firstly, the full wave profile is required for an accurate treatment of a multivalued solution. This can be addressed when solving the flow along the outgoing characteristics by reconstructing the pressure or density profile in space. Secondly, however, the multivalued solution should also be appropriately resolved in space. Assuming that we follow the discrete pieces of information that originate at the gas-liquid interface along the outgoing characteristic, e.g.~as described in Section \ref{sec:numerics_flow}, the spatial resolution of the wave profile is directly dependent on the time step applied to solve the bubble dynamics.} 

While the application of the {rule of equal areas} provides an accurate means to rectify multivalued solutions, it is usually applied as a post-processing step. The clipping method of \citet{Denner2023} can treat multivalued solutions at run time, but does not conserve mass and tends to overestimate the attenuation at shock fronts. To remedy this shortcoming, a run-time shock treatment based on a simple averaging procedure of the multivalued solution is proposed, as illustrated in Figure \ref{fig:shock_averaging}. With the overtaking information parcel $i+1$ tagged to be deleted, the location $r_i$ and invariant $g_i$ of information parcel $i$ are redefined as
\begin{eqnarray}
    r_{i,j} &=& \frac{r_{i+1,j} + r_{i,j}}{2} \label{eq:r_shock_new} \\
    g_{i} &=& \frac{g_{i+1} + g_{i}}{2}. \label{eq:g_shock_new}
\end{eqnarray}
The average of the invariant $g$ follows from the conservation of specific kinetic enthalpy and presumes that the applied time step is sufficiently small so that the invariant $g$ varies linearly in the close vicinity of the shock front. This leaves the enthalpy $h_i$ and the velocity $u_i$ undefined. Considering that the velocity has two solutions that satisfy a given $g$, the robust choice is to average the velocity as 
\begin{equation}
    u_{i,j} = \frac{u_{i+1,j} + u_{i,j}}{2}.  \label{eq:u_shock_new}
\end{equation}
The specific enthalpy $h_i$ is then readily defined by Eq.~\eqref{eq:h_KB_disc}. Furthermore, preliminary tests showed that the choice of quantities that are averaged (i.e., $g_i$ and $u_i$, $h_i$ and $u_i$, or $\rho_i$ and $u_i$) has a negligible impact on the results. After the quantities associated with information parcel $i$ have been redefined, the overtaking information parcel $i+1$ is discarded.

As a first test case, we consider the shock formation of a planar wave emitted in an inviscid fluid by an oscillating planar emitter ($\alpha=0$). The oscillating emitter generates an harmonic acoustic wave that gradually steepens to eventually form a shock front, at which point it starts to develop into a sawtooth wave that decays due to the attenuation acting at the shock front. For such a sawtooth wave, \citet{Fay1931} derived an analytical solution for the pressure amplitude $\Delta p$, given in function of the one-dimensional coordinate $r$ as 
\begin{equation}
    \Delta p (r) = \pm \frac{\pi r_\text{sh}}{r + r_\text{sh}} \, \Delta p_\text{a},
    \label{eq:fay}
\end{equation}
where the shock forming distance of a plane harmonic wave is \citep{Blackstock1966} 
\begin{equation}
    r_\text{sh} = \frac{\rho_0 c_0^3}{2 \pi \beta f_\text{a} \Delta p_\text{a}},
    \label{eq:shock_distance}
\end{equation}
{where $f_\text{a}$ and $\Delta p_\text{a}$ are the initial frequency and pressure amplitude of the emitted acoustic wave. The nonlinearity coefficient $\beta$ for an ideal gas or a Tait fluid is defined as \citep{Hamilton1988,Lauterborn2007a} $\beta = (\gamma+1)/2$.} Note that the solution of Fay, Eq.~\eqref{eq:fay}, is only applicable to ``perfect'' sawtooth waves \citep{Blackstock1966} for $r \gg r_\text{sh}$ and that the excitation amplitude $\Delta p_\text{a}$ ought to be sufficiently small such that the nonlinear dependency of the fluid properties on the pressure (see Section \ref{sec:eos}) is negligible. Figure \ref{fig:Emitter_planar} shows the results obtained using Eqs.~\eqref{eq:drdt_char_disc}-\eqref{eq:h_KB_disc} in conjunction with the clipping-based shock treatment of \citet{Denner2023} and the averaging-based shock treatment proposed above, for an initially harmonic wave in water modeled using the modified Tait EoS ($\beta = 4.075$). At $r=r_\text{sh}+R_0$, the wave profile forms vertical fronts, where $\partial p/\partial r \rightarrow \infty$, as predicted by the theory of \citet{Blackstock1966}. Subsequently, the wave evolves into a fully developed sawtooth wave and starts to decay, with its amplitude asymptotically approaching the solution of \citet{Fay1931}. The clipping procedure of \citet{Denner2023}, whereby a multivalued solution is treated by clipping the ``overtaking'' part of the wave profile, fails to predict this sawtooth wave correctly; only the wave peaks reduce, whereas the wave troughs remain unchanged. This shortcoming is evidently less of an issue for the pressure pulses emitted by a strong bubble collapse for which the clipping procedure has previously been tested \citep{Denner2023}. The proposed averaging procedure is able to predict the decay of both the peaks and troughs of the sawtooth wave accurately.

\begin{figure*}
    \includegraphics[width=0.566666\linewidth]{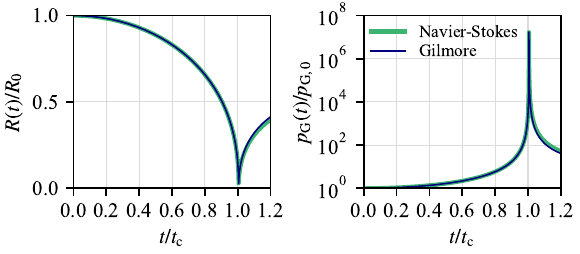}
    \caption{Radius $R(t)$, normalized by the initial radius $R_0$, and gas pressure $p_\text{G}(t)$, normalized by the initial gas pressure $p_\text{G,0}$, as a function of time $t$, normalized by the Rayleigh collapse time $t_\text{c}$, of the Rayleigh collapse of a spherical gas bubble in water, obtained with a Navier-Stokes solver and the Gilmore equation.}
    \label{fig:RP-shock_bubble}
\end{figure*}
\begin{figure*}
    \includegraphics[width=0.85\linewidth]{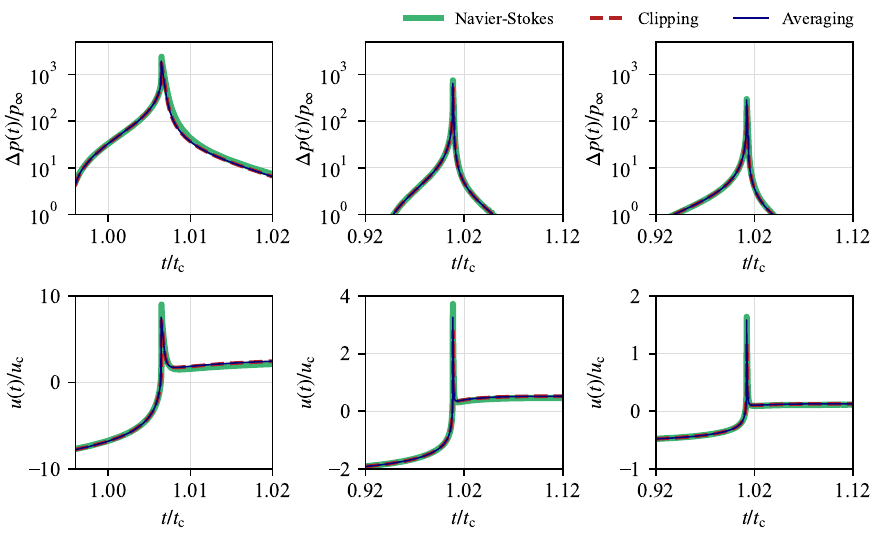}
    \caption{Temporal evolution of the pressure amplitude $\Delta p(r,t)=p(r,t)-p_\infty$, normalized by the constant ambient pressure $p_\infty$, and flow velocity $u(r,t)$, normalized by the characteristic collapse velocity $u_\text{c}=R_0/t_\text{c}$, emitted by the Rayleigh collapse of a spherical gas bubble in water at different radial locations $r = \{ 0.2, 0.5, 1 \} R_0$, obtained with the KBH model described in Section \ref{sec:numerics_flow}. Time $t$ is normalized by the Rayleigh collapse time $t_\text{c}$. The multivalued solutions at the shock front are treated either by clipping the wave profile or by applying the proposed averaging procedure. Results obtained with a Navier-Stokes solver are shown for reference.}
    \label{fig:RP-shock_emissions}
\end{figure*}

\begin{figure*}
  \includegraphics[width=0.85\linewidth]{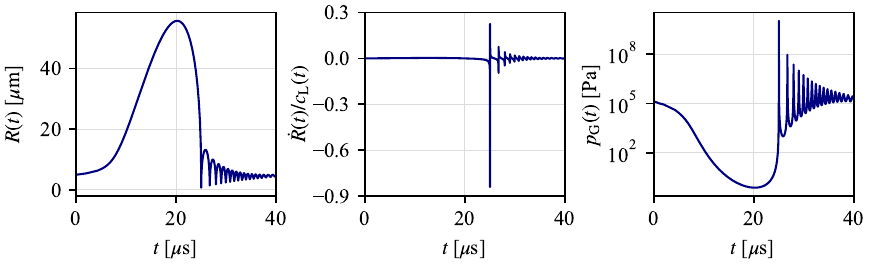}
  \caption{Temporal evolution of the radius $R(t)$, bubble wall Mach number $\dot{R}(t)/c_\text{L}(t)$, and gas pressure $p_\text{G}(t)$ of the sonoluminescence bubble of \citet{Holzfuss2010}, obtained with the Gilmore model.}
  \label{fig:Sonolum_bubble}
\end{figure*}
\begin{figure*}
  \includegraphics[width=0.85\linewidth]{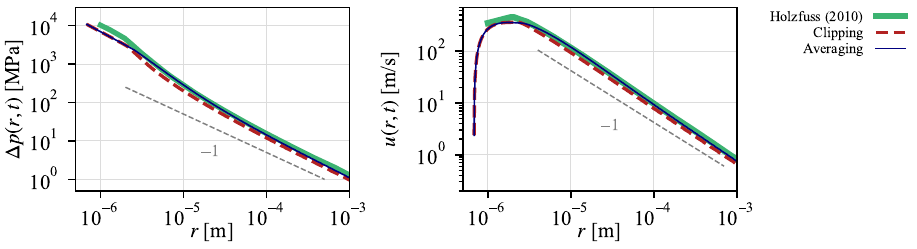}
  \caption{Spatial evolution of the pressure amplitude $\Delta p(r,t)=p(r,t)-p_\infty(t)$ and the flow velocity $u(r,t)$ of the acoustic wave emitted by the primary collapse of the sonoluminescence bubble of \citet{Holzfuss2010}, obtained with the KBH model described in Section \ref{sec:numerics_flow}. The multivalued solutions at the shock front are treated either by clipping the wave profile or by applying the proposed averaging procedure. {The expected pressure and velocity decay in the far field, proportional to $r^{-1}$, is indicated by the gray dashed lines.}}
  \label{fig:Sonolum_minmax}
\end{figure*}

To assess the capabilities of the run-time shock treatment in a more representative scenario, the Rayleigh collapse of a spherical gas bubble ($\alpha=2$) in water is simulated. The bubble has an initial radius $R_0$ and its collapse is driven by an initial gas pressure of $p_\mathrm{G,0} = p_\infty/1000$ compared to a constant ambient pressure of $p_\infty =  10^5 \, \mathrm{Pa}$. The gas is modeled using the NA EoS, see Eqs.~\eqref{eq:pG_NA} and \eqref{eq:rhoG_NA}, with $\rho_{\mathrm{G},0} = 1.2 \, \mathrm{kg/m}^3$ at $p_\infty$, $\kappa=1.4$, and $b=1.5 \times 10^{-3} \, \mathrm{m}^3/\mathrm{kg}$, whereas water is modeled using the modified Tait EoS, with $n=7.15$, $B = 3.046 \, \times 10^8 \, \text{Pa}$, $\rho_0 = 997 \, \text{kg/m}^3$ at $p_\infty$, and $\mu =0.001 \, \mathrm{Pa \ s}$. The surface tension of the gas-liquid interface is neglected. For validation purposes, the results obtained based on the Kirkwood-Bethe hypothesis are compared to a fully resolved simulation using a state-of-the-art Navier-Stokes solver \citep{Denner2018b,Denner2020a}. For this simulation, the pressure in the liquid is initialized as $p(r) = p_\infty + R_0 (p_\mathrm{G,0} - p_\infty) / r$, the bubble is resolved with $1000$ mesh cells per initial bubble radius, and the time step applied at the time when the bubble reaches its minimum radius is $\Delta t \simeq  5 \times 10^{-8} t_\mathrm{c}$, where $t_\mathrm{c} \simeq 0.915 \, R_0 \sqrt{\rho_\infty /p_\infty}$ is the Rayleigh collapse time \citep{Rayleigh1917}. Figure \ref{fig:RP-shock_bubble} shows a close agreement of the bubble radius and gas pressure obtained with the Navier-Stokes solver and with the Gilmore equation, Eq.~\eqref{eq:bubbleODE_gilmore}. The temporal evolution of the pressure and velocity in the vicinity of the bubble is shown in Figure \ref{fig:RP-shock_emissions}, at $r = \{ 0.2, 0.5, 1 \} R_0$. The emitted acoustic wave forms a shock front, with a vertical left slope of the pressure and velocity profiles. Note that in the temporal-evolution plots shown in Figure \ref{fig:RP-shock_emissions}, the left-hand side of the profile reaches the measurement location first. The results obtained with both methods to treat the shock front, either clipping the wave profile or applying the proposed averaging procedure, yield a good agreement with the Navier-Stokes solution for this specific test case. However, clipping the wave profile overpredicts the attenuation of the shock front in the far field, which manifests in a smaller peak pressure and velocity at $r=R_0$.

\begin{figure*}
    \includegraphics[width=0.85\linewidth]{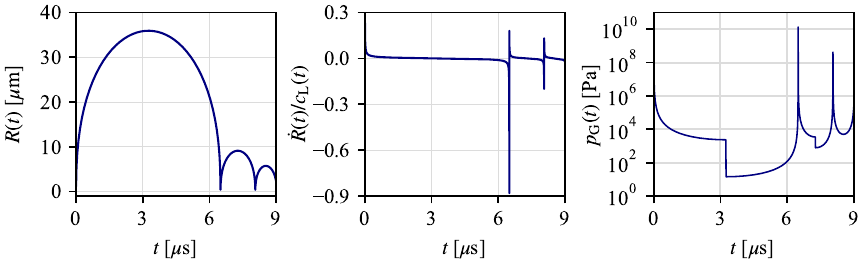}
    \caption{Temporal evolution of the radius $R(t)$, bubble wall Mach number $\dot{R}(t)/c_\text{L}(t)$, and gas pressure $p_\text{G}(t)$ of the laser-induced bubble of \citet{Liang2022}, obtained with the Gilmore model.}
    \label{fig:LIC_bubble}
\end{figure*}
\begin{figure*}
    \includegraphics[width=0.85\linewidth]{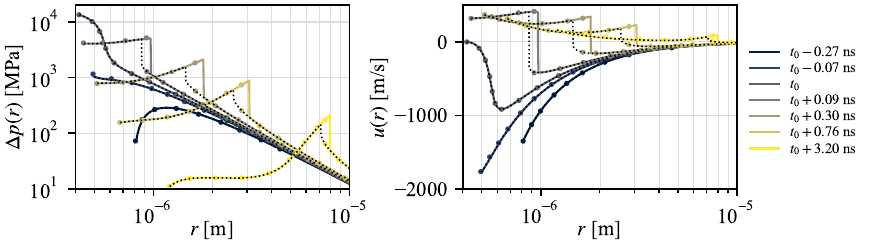}
    \caption{Wave profiles of the pressure amplitude $\Delta p(r,t)=p(r,t)-p_\infty(t)$ and the flow velocity $u(r,t)$ at selected time instances of the acoustic wave emitted by the first collapse of the laser-induced bubble of \citet{Liang2022}, obtained with the KBH model described in Section \ref{sec:numerics_flow}. The solid line shows the results obtained by applying the proposed averaging procedure at the shock front, the dotted line shows the results obtained by clipping the wave profile, and the colored dots show the corresponding results of \citet{Liang2022}. Time $t_0$ is the time at which the peak pressure is emitted.}
    \label{fig:LIC_wave}
\end{figure*}

The strong collapse and shock formation of an argon bubble that produces so\-no\-lu\-mi\-nes\-cence \citep{Holzfuss2010} and of a laser-induced bubble \citep{Liang2022} have previously been studied with respect to the emission of shock waves using KBH models. In both studies \citep{Holzfuss2010,Liang2022}, the rule of equal areas was applied to rectify the multivalued solutions at the shock front. The sonoluminescence bubble considered by \citet{Holzfuss2010} has an initial radius of $R_0=5 \, \mu \text{m}$ and is excited with a frequency of $f_\text{a}=23.5 \, \text{kHz}$ and a pressure amplitude of $\Delta p_\text{a} = 145 \, \text{kPa}$, whereas the laser-induced bubble studied by \citet{Liang2022} is generated by a laser pulse with $20 \, \mu \text{J}$ energy and $265 \, \text{fs}$ duration, and reaches a maximum bubble radius of approximately $36 \, \mu \text{m}$. In both cases, water is modeled by the modified Tait EoS. The exact parameters of these cases can be found in the respective publications, or the recent work of \citet{Denner2023}. For orientation, the radius $R$, bubble wall Mach number $\dot{R}/c_\text{L}$, and gas pressure $p_\text{G}$ of the sonoluminescence and laser-induced bubbles are shown as a function of time in Figures \ref{fig:Sonolum_bubble} and \ref{fig:LIC_bubble}, respectively. Note that the discontinuous changes in gas pressure of the laser-induced bubble are intentional and account for the complex mass-transfer dynamics of this bubble, as proposed by \citet{Liang2022}. In Figure \ref{fig:Sonolum_minmax}, the pressure and velocity amplitude of the acoustic wave emitted by the primary collapse of the sonoluminescence bubble are shown, where subtle differences can be observed dependent on the treatment of the developing shock front. The proposed averaging procedure exhibits a tendency to larger pressure and velocity values compared to simply clipping the wave profile. Nevertheless, the results obtained with both methods to treat the shock front at run time compare very well with the results of \citet{Holzfuss2010}. The spatial profiles of the acoustic wave produced by the collapse of the laser-induced bubble, see Figure \ref{fig:LIC_wave}, exhibit a comparable trend: the proposed averaging procedure yields a faster shock front with larger pressure and velocity values compared to both the results of \citet{Liang2022} and the results obtained by clipping the wave profile. The observed differences are confined to the immediate vicinity of the shock front and are largest in the region in which the shock dissipation is strongest.

\subsection{Complex impedance of curved acoustic waves}
\label{sec:results_impedance}

The acoustic impedance experienced by a curved acoustic wave is complex, see Eq.~\eqref{eq:impedance}, and depends on the radius of curvature of the acoustic wave relative to its wavenumber. Only for relatively short waves ($kr \gg 1$) does the simple definition $z=\rho c$ of the specific acoustic impedance hold in good approximation.

As discussed in Section \ref{sec:liquid_expl}, the explicit velocity expression given by Eq.~\eqref{eq:u_expl_expanded} is the sum of the displacement of the fluid by the moving gas-liquid interface and the acoustic particle velocity associated with the emitted acoustic wave. According to Eq.~\eqref{eq:u_expl_expanded}, the acoustic particle velocity $u_1$ is, thus, given as
\begin{equation}
   u_1(r,t) = \left[\frac{R(\tau)}{r(t)}\right]^{\alpha/2} \frac{h_\text{L}(\tau)-h_\infty(\tau)}{c(r,t)+u(r,t)} .
\end{equation}
Applying the isentropic relation $\text{d}h = \text{d}p/\rho$ and assuming the acoustic wave has a small pressure amplitude, such that $u\ll c$, $\rho=\rho(r,t)=\rho_\text{L}(\tau)$ and $c=c(r,t)=c_\text{L}(\tau)$, the velocity is
\begin{equation}
    u_1(r,t) = \left[\frac{R(\tau)}{r(t)}\right]^{\alpha/2} \frac{p_\text{L}(\tau)-p_\infty(\tau)}{\rho c},
\end{equation}
or, using the notation used in the acoustic expansions, with $p = p_0 + p_1$, 
\begin{equation}
    u_1(r,t) = \left[\frac{R(\tau)}{r(t)}\right]^{\alpha/2} \frac{p_1(\tau)}{\rho c}.
\end{equation}
This expression is evidently not dependent on the relative curvature of the emitted wave and, therefore, suggests that velocity expressions derived from the Kirkwood-Bethe hypothesis do not account for the complex acoustic impedance of curved waves.

\begin{figure*}[t]
  \includegraphics[width=0.85\linewidth]{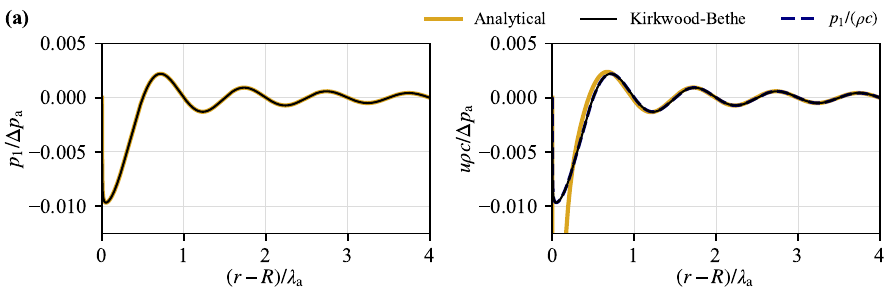}
  \includegraphics[width=0.85\linewidth]{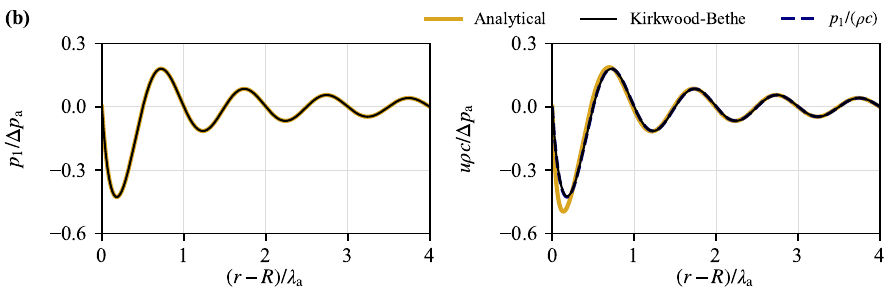}
  \includegraphics[width=0.85\linewidth]{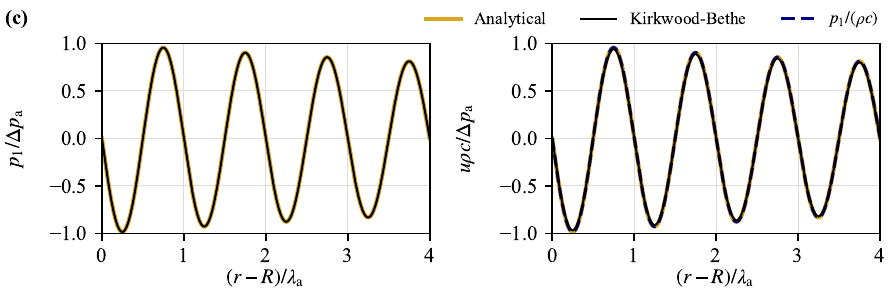}
  \caption{The pressure (left) and velocity (right) of acoustic waves emitted by a pulsating spherical emitter with (a) $k_\text{a} R_0 = 0.01$, (b) $k_\text{a} R_0 = 1$, and (c) $k_\text{a} R_0 = 100$, predicted by the Kirkwood-Bethe hypothesis. The analytical solution for curved acoustic waves, Eqs.~\eqref{eq:p1_kr} and \eqref{eq:u_kr}, and the particle velocity for plane acoustic waves are shown for reference.}
  \label{fig:impedance}
\end{figure*}

To test this, we consider a spherical emitter situated in water that oscillates and emits acoustic waves at angular frequency $\omega_\text{a}$, chosen such that a specific dimensionless wavenumber $k_\text{a}R_0$ is obtained, with $k_\text{a}= \omega_\text{a}/c$. The pressure at the emitter wall and the time-dependent velocity of the emitter wall are defined as 
\begin{eqnarray}
    p_\text{L}(t) &=& p_0 + \Delta p_\text{a} \, \cos\left(\omega_\text{a} t - \frac{\pi}{2}\right)\\
    \dot{R}(t) &=& \frac{\Delta p_\text{a}}{\rho c} \, \cos\left(\omega_\text{a} t - \frac{\pi}{2}\right),
\end{eqnarray}
respectively. The considered pressure amplitude $\Delta p_\text{a}$ is sufficiently small such that the pressure-dependency of the liquid density and speed of sound are negligible. At the emitter wall, pressure and velocity are, therefore, in phase and there is no relative velocity between the emitter wall and the fluid. The water surrounding the emitter is described using the modified Tait EoS, with $n=7.15$, $B=3.046 \times 10^8 \, \text{Pa}$, and $\rho_0 = 997 \, \text{kg/m}^3$. Following Eq.~\eqref{eq:impedance}, the analytical solutions for the acoustic pressure and particle velocity are given as\citep{Kinsler2000}
\begin{equation}
    p_1(r,t) = \frac{R}{r} \, \Delta p_\text{a} \, \cos\left(\omega_\text{a} t - \frac{\pi}{2}- k (r-R) \right) \label{eq:p1_kr}
\end{equation}
\begin{multline}
  u_1(r,t) = \frac{R}{r} \, \frac{\Delta p_\text{a}}{\rho c \cos(\theta)} \, \cos\left(\omega_\text{a} t - \frac{\pi}{2} - k (r-R) - \theta \right), \label{eq:u1_kr}
\end{multline}
respectively, where the complex impedance manifests in the phase shift $\theta = \pi/2 - \arctan(kr)$. Taking the oscillating emitter into account by inserting the analytical particle velocity defined by Eq.~\eqref{eq:u1_kr} into the flow velocity given by Eq.~\eqref{eq:u_expl}, the analytical solution for the flow velocity is
\begin{equation}
    u(r,t) = \frac{R^2}{r^2} \dot{R}(t) - \frac{R}{r} u_1(r,t) + u_1(r,t).\label{eq:u_kr}
\end{equation}

Figure \ref{fig:impedance} shows the spatial profiles of the pressure and velocity generated by the spherical emitter ($\alpha=2$) for $k_\text{a} R_0 \in \{0.01,1,100\}$, predicted by the Kirkwood-Bethe hypothesis, using Eqs.~\eqref{eq:drdt_char_disc}-\eqref{eq:h_KB_disc}. For all three cases, the pressure predicted by the Kirkwood-Bethe hypothesis is in virtually perfect agreement with the analytical solution, Eq.~\eqref{eq:p1_kr}. However, the velocity profile predicted by the Kirkwood-Bethe hypothesis agrees with the analytical solution for a curved wave, Eq.~\eqref{eq:u_kr}, only for the case with $k_\text{a} R_0 = 10^2$, i.e.~for a short wave with $kr \gg 1$. As the dimensionless wavenumber decreases, i.e.~longer waves, the accuracy of the velocity predicted by the Kirkwood-Bethe hypothesis deteriorates near the emitter. Despite this shortcoming, the Kirkwood-Bethe hypothesis predicts the velocity accurately after the wave has traveled for more than approximately one wavelength ${\lambda_\text{a}}$, which corresponds to $k_\text{a}(r-R_0) \gtrsim 2\pi$. We can, therefore, conclude that the Kirkwood-Bethe hypothesis does not account for the complex acoustic impedance of curved waves and, as a consequence, fails to predict the particle velocity of long waves ($kr \lesssim 1$) close to the emitter correctly. In fact, the velocity ODE arising from the Kirkwood-Bethe hypothesis, Eq.~\eqref{eq:dudt_char_disc}, predicts $u_1 \simeq p_1/(\rho c)$ for all three cases, as observed in Figure \ref{fig:impedance}, irrespective of the dimensionless wavenumber $kr$. Nevertheless, the predictions provided by the Kirkwood-Bethe hypothesis are accurate for $r \gtrsim R+\lambda_\text{a}$.

Recent experimental and numerical studies that reported accurate predictions of KBH models stand in apparent contradiction to the shortcoming associated with the complex nature of the acoustic impedance. For instance, the experimental measurements of the pressure and velocity of shock waves emitted by laser-induced cavitation bubbles of \citet{Lai2022} and the fully resolved numerical results of the Rayleigh collapse of a spherical gas bubble shown in Figure \ref{fig:RP-shock_emissions} are in very good agreement with the corresponding predictions provided on the basis of the Kirkwood-Bethe hypothesis. However, in these cases, which are representative for pressure-driven bubble dynamics, including underwater explosions as well as medical and sonochemical applications, the length of the emitted pressure pulse is short, with $kR = \mathcal{O}(1)$ or larger \cite{Lai2022, Denner2023, Wen2023a, Liang2022}. In addition, the velocity very close to the bubble is dominated by the motion of the gas-liquid interface and even though the particle velocity is nominally large, it is typically less than $10\%$ of the total propagation velocity $c+u$.

\section{Conclusions}
\label{sec:conclusions}

Originally developed to estimate the peak pressure of an underwater explosion at some distance, the Kirkwood-Bethe hypothesis has given rise to a collection of models that describe pressure-driven bubble dynamics, cavitation and underwater explosions, as well as their acoustic emissions, with remarkable accuracy. By using standard numerical integration methods and modern computational equipment, these models allow to predict complex processes of fluid dynamics and nonlinear acoustics, such as the laser-induced cavitation bubble discussed in Section \ref{sec:results_shocks}, in a fraction of a second. Here, the theoretical foundations of the Kirkwood-Bethe hypothesis and contemporary models derived from it have been reviewed, as well as generalized from the typically considered spherical symmetry, to account for spherically symmetric, cylindrically symmetric, and planar one-dimensional domains. In addition, a new method to treat multivalued solutions associated with the formation of shock fronts at simulation run time has been proposed, which further improves the predictive capabilities of the models based on the Kirkwood-Bethe hypothesis. Even though the Kirkwood-Bethe hypothesis fails to account for the curvature of acoustic waves, as they are, for instance, emitted by a bubble collapse, it has been shown to provide accurate predictions under the specific conditions associated with pressure-driven bubble dynamics, cavitation and underwater explosions. 

Despite the extensive and formidable work achieved by the research community over more than eight decades, models based on the Kirkwood-Bethe hypothesis still lack some basic capabilities that can further extend their applicability and improve their accuracy. For instance, although the Kirkwood-Bethe hypothesis can produce accurate pressure and velocity predictions under realistic scenarios, it does not account for heat transfer. As a first step, the recent combination of the Kirkwood-Bethe hypothesis with the Noble-Abel stiffened-gas equation of state for bubble dynamics \citep{Denner2021} and acoustic emissions \citep{Denner2023} provides a consistent and accurate basis for predicting the temperature in the liquid surrounding a gas bubble or cavity. While adiabatic temperature changes can be readily predicted in this way, thermal conduction and viscous heating, such as the heat introduced by the energy dissipation at a shock front \citep{Liang2022}, are not accounted for. In this respect, medical ultrasound stands out as a field that requires high-fidelity temperature predictions if computational tools are to play a more prominent role in the design of new treatments, since temperature changes of only $1 \, \text{K}$ can be safety relevant \citep{terHaar2011}. More generally, bubbles are neither stationary nor do they appear in isolation or in a liquid of quasi-infinite {extent} in most engineering applications, which contradicts some of the main assumptions underpinning the available models based on the Kirkwood-Bethe hypothesis. Recent advances for methods developed based on the (quasi-)acoustic assumption point to possible solutions that may be incorporated into models based on the Kirkwood-Bethe hypothesis, to account for the forces on bubbles as they move through a liquid \citep{Zhang2023b}, the acoustic interaction of multiple bubbles \citep{Fan2021}, and the influence of confinement and nearby walls \citep{Fu2023}.

\section*{Conflict of interest}
\noindent
The author has no conflicts that could have influenced this work. 

\section*{Data availability}
\noindent
All presented results were produced, and can be readily reproduced, with the open-source software library {\tt APECSS}, as referenced in Section \ref{sec:results}.

\section*{References}

\end{document}